\PassOptionsToPackage{usenames,dvipsnames}{xcolor}
\documentclass[sigconf, nonacm]{acmart}

\usepackage{graphicx}
\usepackage{enumitem}
\usepackage{subcaption}
\usepackage{booktabs} 
\usepackage{mathtools}
\usepackage{xspace}
\usepackage{appendix}
\usepackage{subfiles} 

\DeclarePairedDelimiter\floor{\lfloor}{\rfloor}
\usepackage{tikz}

\BeforeBeginEnvironment{appendices}{\clearpage}

\newcommand\vldbdoi{10.14778/3494124.3494141}

\newcommand\vldbavailabilityurl{https://github.com/baotonglu/apex}
\newcommand\vldbpagestyle{empty} 
\newcommand*\circled[1]{\tikz[baseline=(char.base)]{
            \node[shape=circle,fill=.,inner sep=0pt] (char) {\color{-.}\textsf\footnotesize #1};}}

\newcommand{\bptree}{B+-tree\xspace}
\newcommand{\lbtree}{LB+Tree\xspace}
\newcommand{\dptree}{DPTree\xspace}
\newcommand{\utree}{uTree\xspace}
\newcommand{\fptree}{FPTree\xspace}
\newcommand{\longlat}{\texttt{Longlat}\xspace}
\newcommand{\longitudes}{\texttt{Longitudes}\xspace}
\newcommand{\ycsb}{\texttt{YCSB}\xspace}
\newcommand{\tpce}{\texttt{TPC-E}\xspace}
\newcommand{\lognormal}{\texttt{Lognormal}\xspace}
\newcommand{\fb}{\texttt{FB}\xspace}

\definecolor{comment-red}{rgb}{1,0,0}

\usepackage{fix-cm}
\usepackage[colorinlistoftodos,prependcaption,textsize=small]{todonotes}
\usepackage{varwidth}

\def\thepaperkeywords{Learned index, persistent memory, Optane DCPMM}
\def\thepapertitle{APEX: A High-Performance Learned Index on Persistent Memory (Extended Version)}

\usepackage{float}
\hypersetup{
  pageanchor=true,
  plainpages=false,
  pdfpagelabels,
  bookmarks,
  bookmarksnumbered,
  pdfborder=0 0 0,  
  colorlinks=true,
  linkcolor=ACMRed,
  anchorcolor=ACMRed,
  citecolor=Blue, 
  pdftitle={\thepapertitle},
  pdfauthor={},
  pdfsubject={},
  pdfkeywords={\thepaperkeywords},
}

\usepackage{hyperref}

\usepackage{breakurl}
\begin{document}

\renewcommand\thefigure{\Alph{figure}}    
\renewcommand\thetable{\Alph{table}}    

\title{\thepapertitle}

\author{Baotong Lu}
\affiliation{%
  \institution{The Chinese University of Hong Kong}
}
\email{btlu@cse.cuhk.edu.hk}

\author{Jialin Ding}
\affiliation{%
\institution{Massachusetts Institute of Technology}
 }
\email{jialind@mit.edu}

\author{Eric Lo}
\affiliation{%
  \institution{The Chinese University of Hong Kong}
}
\email{ericlo@cse.cuhk.edu.hk}

\author{Umar Farooq Minhas}
\affiliation{%
  \institution{Microsoft Research}
}
\email{ufminhas@microsoft.com}

\author{Tianzheng Wang}
\affiliation{%
  \institution{Simon Fraser University}
}
\email{tzwang@sfu.ca}


\begin{abstract}
The recently released persistent memory (PM) offers high performance, persistence, and is cheaper than DRAM. 
This opens up new possibilities for indexes that operate and persist data directly on the memory bus. 
Recent learned indexes exploit data distribution and have shown great potential for some workloads. 
However, none support persistence or instant recovery, and existing PM-based indexes typically evolve B+-trees without considering learned indexes.  

This paper proposes APEX, a new PM-optimized learned index that offers high performance, persistence, concurrency, and instant recovery. 
APEX is based on ALEX, a state-of-the-art updatable learned index, to combine and adapt the best of past PM optimizations and learned indexes, allowing it to reduce PM accesses while still exploiting machine learning. 
Our evaluation on Intel DCPMM shows that APEX can perform up to $\sim$15$\times$ better than existing PM indexes and can recover from failures in $\sim$42ms. 
\end{abstract}

\maketitle

\pagestyle{\vldbpagestyle}
\begingroup
\renewcommand\thefootnote{}\footnote{\noindent
This document is an extended version of "APEX: A High-Performance Learned Index on Persistent Memory", which will appear in The 48th International Conference on Very Large Data Bases (VLDB 2022). This document is freely provided under Creative Commons.
}\addtocounter{footnote}{-1}\endgroup

\renewcommand\thetable{\arabic{table}}    
\renewcommand\thefigure{\arabic{figure}}    
\setcounter{figure}{0}   
\setcounter{table}{0}   
\setcounter{section}{0}
\setcounter{page}{1}

\section{Introduction}
\label{sec:intro}
Modern data systems use fast memory-optimized indexes (e.g.,
B+-trees)~\cite{BwTree,Masstree,ART,HOT,CSBTree} for high
performance. As data size grows, however, scalability is limited by
DRAM's high cost and low capacity: OLTP indexes alone can occupy $>$ 55\%
of total memory~\cite{HybridIndex}.
Byte-addressable persistent memory (PM)~\cite{Intel3DXP,Memristor,PCM}
offers persistence, high capacity, and lower cost compared to DRAM.
The recently released Intel Optane DCPMM~\cite{DCPMM} is available in
128--512GB DIMMs, yet 128GB DRAM DIMMs are rare and priced
$\sim$5--7$\times$ higher than 128GB DCPMM~\cite{DCPMM-DRAM-Price}.  
Although PM is more expensive than SSDs, it offers better performance, making it an attractive option to complement limited/expensive DRAM.
These features have led to numerous
PM-optimized
indexes~\cite{Venkataraman2011,WORT,ROART,FPTree,Dash,BzTree,Hwang2018,lbtree,Chen2015,dptree,CCEH,LevelHashing,NV-Tree,PFHT,utree}
that directly persist and operate on PM.  Some also 
support instant recovery to reduce down time.

\begin{figure}[t]
	\centering
  \includegraphics[width=0.49\columnwidth]{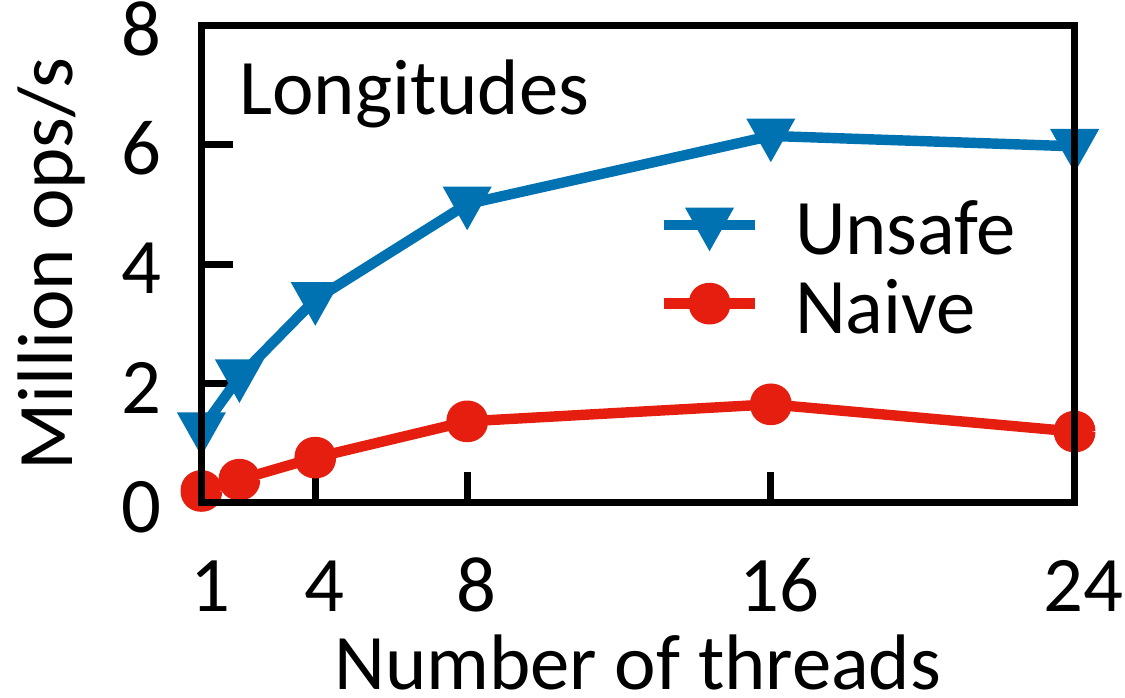}
  \hfill
  \includegraphics[width=0.49\columnwidth]{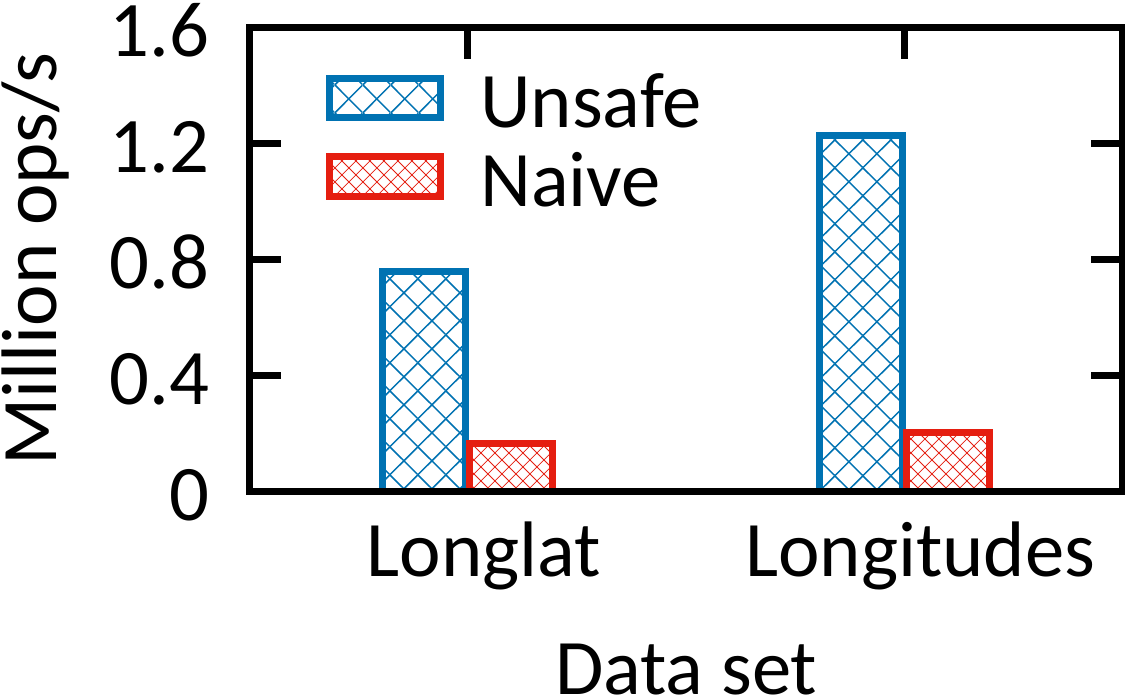}
  \caption{
    Insert scalability (left) and single-thread throughput (right) of ALEX~\cite{ALEX} on PM. 
    Naively using PMDK (\textsf{Naive}) limits performance due to PM's limited bandwidth.
    Directly running it on PM (\textsf{Unsafe}) further loses crash consistency.} 
	\label{fig:alex-txn}
\end{figure}

Most (if not all) existing PM indexes are based on
B+-trees or hash tables which are agnostic to data distribution. 
As demonstrated by recent learned
indexes~\cite{LearnedIndex,FITing-Tree,ALEX,RadixSpline,PGM,ZM,ML-Index,XIndex,LISA,RSMI,qd-tree,Tsunami,Flood},
indexes can be implemented as machine learning models that
predict the location of target values given a search key.  
Suppose all values are stored in an array sorted by key
order, a linear model trained from the data can directly
output the value's position in the array. If 
keys are continuous integers (e.g., 0--100 million), the value
mapped to key $k$ can be accessed by \texttt{array[k]}.  Such
model-based search gives $O(1)$ complexity and the entire index
is as simple as a linear function.
Some learned indexes (e.g., ALEX~\cite{ALEX}) also support updates and inserts.  
They typically use a hierarchy of
models~\cite{ALEX,LearnedIndex,PGM} that form a tree-like structure to
improve accuracy. 
However, individual nodes could be much bigger (e.g., 16MB in ALEX~\cite{ALEX}), leading to very high
fanout (e.g., 2$^{16}$) and low tree depth (e.g.,
2), making search operations lightweight even for very large data
sizes.

\subsection{When Learned Indexing Meets Persistent Memory: The Old Tricks No Longer Work!}
\label{sec:learn-pm}
We observe learned indexing is a natural fit for PM: Real PM (Optane
DCPMM) exhibits $\sim$3--14$\times$ lower bandwidth than DRAM \cite{UCSDGuide,PiBench},
whereas model-based search is especially good 
at reducing memory accesses.
But learned indexes were designed based on DRAM without considering PM
properties, and prior PM indexes did not leverage machine learning.
It remains challenging for learned indexes to work well on PM. 

\textbf{\it\textsf{Challenge 1: Scalability and Throughput.}}
Although learned indexes are frugal in bandwidth usage for
lookups, they still exhibit excessive PM
accesses for inserts. This is because learned indexes require data
(key-value pairs) be maintained in sorted order, which may require shifting records for inserts.
Figure~\ref{fig:alex-txn} shows its impact by running
ALEX~\cite{ALEX}---a state-of-the-art learned index---on PM without
any optimizations (denoted as \textsf{Unsafe}).\footnote{Original ALEX
does not support concurrency.  On each core we run an ALEX instance
that works on a data partition.}  Since PM exhibits asymmetric read/write
bandwidth with writes being 3--4$\times$ lower, frequent
record shifting can easily exhaust write bandwidth and eventually
limit insert scalability and throughput.  
This problem was confirmed by recent work~\cite{shimin21}. 
Similar issues were also found in B+trees. A common solution is to use
unsorted nodes~\cite{BzTree,FPTree,NV-Tree,lbtree} that accept inserts
in an append-only manner, but require linear search for lookups.  
This is reasonable for small B+-tree nodes (e.g., 256B--1KB), but 
for model-based operations to work well, 
it is critical to use large nodes (e.g., up to 16MB in ALEX) with sorted data.
Structural modification operations (SMOs) 
become more expensive with more PM accesses and higher synchronization cost: typically only one thread can work on a node during an SMO.

\textbf{\it\textsf{Challenge 2: Persistence and Crash Consistency.}}
A key feature of persistent indexes is to ensure correct
recovery across restarts and power cycles.  Prior learned indexes were
mostly based on DRAM and did not consider persistence issues.  
Simply running a learned index on PM does not
guarantee consistency.  Any operations that involve writing more than
eight bytes could result in inconsistencies as 
currently only 8-byte PM writes are atomic.  Although
recent work~\cite{PMDK,RECIPE} provides easy ways to
convert DRAM indexes to work on PM with crash consistency, they
are either not general-purpose, or incur very high overhead. For
example, PMDK~\cite{PMDK}---the de-facto standard PM 
library---allows developers to wrap operations in 
transactions to easily achieve crash consistency.  As shown in
Figure~\ref{fig:alex-txn}, compared to the
\textsf{Unsafe} variant, this approach (\textsf{Naive}) scales poorly with 
low single-thread throughput because it uses 
heavyweight logging which incurs write amplification and
extra persistence overhead, depleting the scarce PM 
bandwidth. 

\subsection{APEX}
This paper presents \textit{APEX}, \underline{a}
\underline{p}ersistent l\underline{e}arned inde\underline{x} that
retains the benefits of learned indexes while guaranteeing crash consistency on PM and supporting instant recovery and scalable concurrency.

APEX is carefully designed to address the challenges. 
\circled{\textsf{1}} We observe a data (leaf) node in a
learned index can be regarded as a hash table where a linear model is
effectively used as an order-preserving hash function.  A collision
results when the model predicts the same position for multiple
keys. Based on this observation, we develop a collision-resolving
mechanism, probe-and-stash, to retain efficient model-based search
while avoiding excessive PM writes. 
APEX also achieves crash consistency for all operations with low
overhead.  \circled{\textsf{2}} To reduce synchronization and SMO overheads, 
APEX adopts variable node sizes with smaller data nodes (256KB) and larger inner nodes 
(up to 16MB) as SMOs on inner nodes are relatively rare. 
The former allows lightweight SMOs; the latter allows 
shallower trees with high fanout. We
also design a lightweight concurrency control protocol 
to reduce synchronization overhead.  \circled{\textsf{3}}
Similar to prior work, APEX stores certain frequently used 
metadata in DRAM to reduce the impact of PM's higher latency and lower
bandwidth.  
Unlike many other PM indexes, however, APEX
does so while providing instant recovery.  The key is to ensure 
DRAM-resident metadata can be re-constructed quickly and most recovery work can be deferred.

Our evaluation using realistic workloads shows that APEX is up
to $\sim$15$\times$ faster as compared to the state-of-the-art PM
indexes~\cite{BzTree,lbtree,Hwang2018,dptree,FPTree,utree} while achieving high
scalability and instant recovery. We made APEX
open-source at {\url{\vldbavailabilityurl}}.

We make four contributions.  First, APEX brings
persistence to learned indexes which is a missing but a necessary
feature~\cite{ML-in-DB}, bringing learned indexing another step closer
to practical adoption.  Second, APEX combines the best of
PM and machine learning (high performance with a small storage footprint). 
Third, we propose a set of techniques to
implement learned indexes on real PM.  APEX is based on ALEX,
but our techniques (e.g., probe-and-stash and judicious use of DRAM) are general-purpose and
applicable to other indexes. 
Last, we provide a comprehensive evaluation and
compare APEX with prior PM indexes to validate our design decisions.

\section{Background and Related Work}
\label{sec:bg}
In this section, we provide background on PM hardware, existing techniques in PM-optimized indexes, and learned indexes.

\subsection{Intel Optane DC Persistent Memory}
\label{subsec:pm}
Among various scalable PM types~\cite{Intel3DXP,Memristor,PCM}, 
only Intel Optane DCPMM based on 3D XPoint
is commercially available; 
so we target it in this paper. 
DCPMM can run in \emph{Memory} or \emph{App Direct}~\cite{UCSDGuide} modes. 
The former leverages PM's high capacity to present bigger but slower volatile memory
with DRAM as a hardware-managed cache. 
The latter allows software to judiciously use DRAM and PM with persistence. 
We leverage PM's persistence using  
the App Direct mode and frugally use DRAM to boost performance.
Both DRAM and PM are behind volatile CPU caches,
and the CPU may reorder writes to PM. For correctness, software
must explicitly issue cacheline flushes  
(\texttt{CLWB}/\texttt{CLFLUSH}) and fences to force data to 
the ADR domain~\cite{IntelManual}, 
which includes a write buffer and a write
pending queue with persistence guarantees upon failures~\cite{UCSDGuide}.  
Once in ADR, not necessarily in PM media, data is considered persisted.

Although writing DCPMM media exhibits higher latency than reads, recent  
work~\cite{UCSDGuide,PMPrimitives} showed
that end-to-end read latency 
is often higher 
as a
write commits once it reaches ADR while PM
reads often require fetching data from raw media unless cached.  
DCPMM exhibits $\sim$300ns random read
latency, $\sim$4$\times$ slower than DRAM's; the end-to-end write
latency can be lower than $\sim$100ns~\cite{UCSDGuide}.
DCPMM
bandwidth is also lower than DRAM's. 
Compared to DRAM, it exhibits 3$\times$/8$\times$ lower sequential/random read bandwidth and 11$\times$/14$\times$
lower sequential/random write bandwidth.  Its write
bandwidth is also 3--4$\times$ lower than read bandwidth.
DCPMM's internal access granularity is 256 bytes (one
``XPLine'')~\cite{UCSDGuide}.  To serve a 64-byte cacheline read,
it internally loads 256B and returns the requested 64B.
DCPMM also writes in 256B units.  Thus, $<256$B accesses lead to read and write amplification
that wastes bandwidth. For high performance, software should consider PM access in 256B units.

\subsection{PM-Optimized Indexing Techniques}
\label{subsec:pmtree}
Numerous PM-optimized indexes~\cite{FPTree,Dash,BzTree,Hwang2018,lbtree,Chen2015,dptree,CCEH,LevelHashing,NV-Tree,ROART,utree} have been proposed based on B+-trees and hash tables. 
They mainly optimize for crash consistency and performance. 
We give an overview of the key techniques proposed by prior PM indexes which APEX adapts for learned indexing in later sections. 

\textbf{Reducing PM Accesses.}
Many PM B+-trees~\cite{FPTree,BzTree,lbtree,Chen2015,NV-Tree} use unsorted leaf nodes to avoid shifting records upon inserts. 
A record can be inserted into any free slot in a node; free space is tracked by a bitmap. 
This reduces PM writes but requires linear scan for point queries. 
To alleviate such cost, FPTree~\cite{FPTree} accompanies each key with a fingerprint (a one-byte hash of the key) to predict if a key possibly exists. 
Lookups then only access records with matching fingerprints, removing unnecessary PM accesses. 
Some hash tables~\cite{Dash,LevelHashing} use additional stash buckets to handle collisions. 
This reduces expensive PM accesses in the main table that would otherwise be necessary (e.g., chaining requires more dynamic PM allocations and linear probing may issue many reads). 
The tradeoff is lookups may need to check stashes in addition to the main table, but this can be largely alleviated using fingerprints for stashes~\cite{Dash}. 

\textbf{Instant Recovery.} 
PM's byte-addressability and persistence allow placing the entire index on the memory bus and recover from failures without much work~\cite{Chen2015,Hwang2018,BzTree,Dash,LevelHashing}, reducing service down time. 
Lazy recovery~\cite{Dash,Hwang2018} is a well-known technique to realize this. 
Here we describe a recent approach~\cite{Dash}. 
The index maintains a global version number $G$ in PM and each PM block (e.g., an inner or leaf node) is associated with a local version number $L$. 
Upon restart, $G$ is incremented by one, after which the system is ready to serve requests. 
Individual nodes are only recovered later by the accessing threads if $G$ and $L$ do not match. 
This way, the ``real'' recovery work is amortized over runtime, in exchange for instant and bounded recovery time (incrementing one integer).

\textbf{(Selective) Persistence.}
To overcome PM's lower performance, some PM indexes~\cite{dptree,lbtree,FPTree} leverage DRAM by placing reconstructable data (e.g., B+-tree inner nodes) in DRAM for fast search. 
Upon restart, the DRAM-resident data must be reconstructed from data in PM, before the system can start to serve requests. 
This is doable for B+-trees using bulk loading algorithms. 
The downside is recovery time scales with data size, sacrificing instant recovery. 

\textbf{Concurrency Control.}
Both lock-free and lock-based designs have been proposed for PM indexes. 
Traditionally, lock-free programming has been difficult on PM. 
Recent work~\cite{PMwCAS,BzTree} has demonstrated the feasibility of building PM-based lock-free indexes more easily, but the overhead is not negligible~\cite{PiBench}.
Traditional node-level locking causes exclusive accesses and incur more PM writes when acquiring/releasing read locks.
So lock-based designs are often combined with lock-free read and/or hardware transactional memory (HTM)~\cite{lbtree,FPTree} to reduce PM writes. 
FPTree~\cite{FPTree} uses HTM for inner nodes and locking for leaf nodes. 
HTM performs well under low contention, but is not robust due to issues like spurious aborts~\cite{PiBench}. 
Some proposals~\cite{dptree,Dash} use optimistic locking that requires 
locking for writes, and reads can proceed without holding a lock but must verify the read data is consistent. 
This is usually done by checking a version number associated with the data item did not change, which if happened, 
would cause the read operation to be retried. 

\begin{figure}[t]
\centering
\includegraphics[width=\columnwidth]{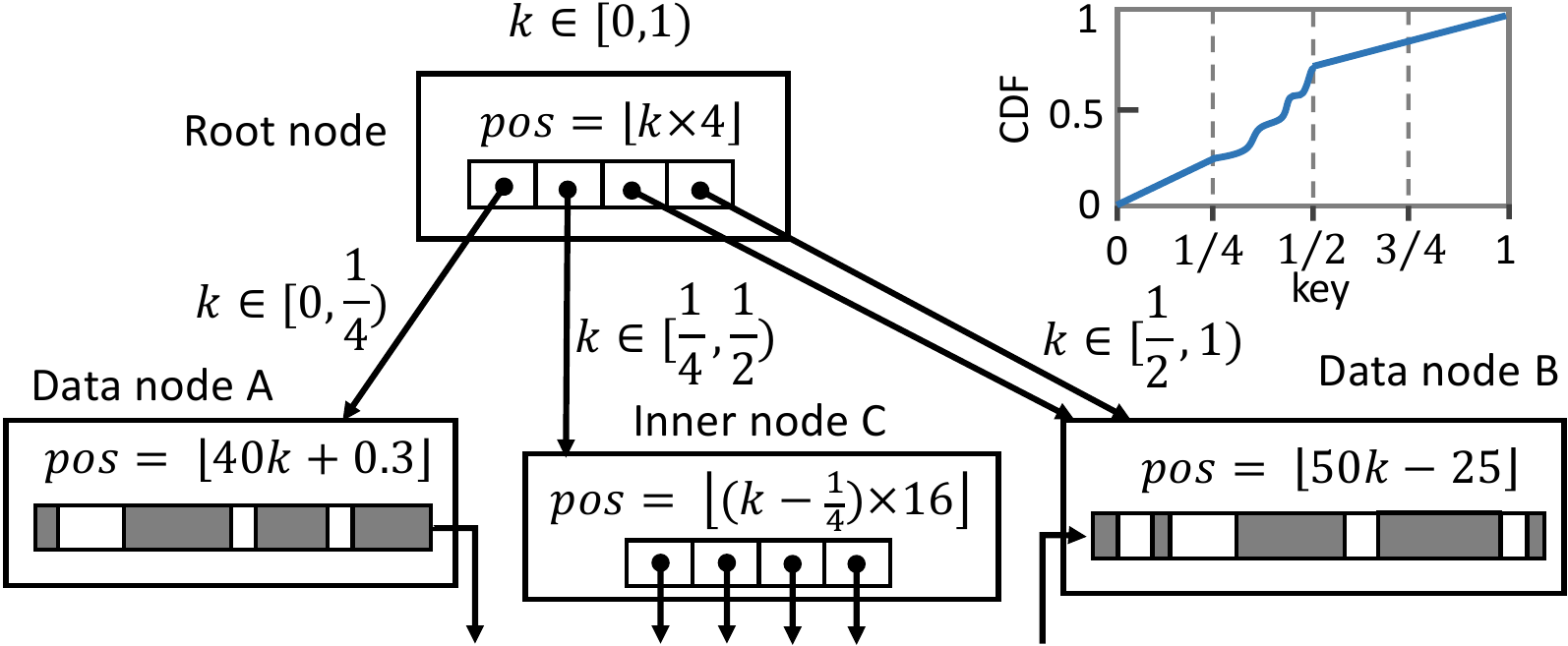}
\caption{ALEX structure: in addition to models, inner/leaf (data) nodes store pointers/records in gapped arrays~\cite{ALEX}.
}
\label{fig:alex} 
\end{figure}

\subsection{Learned Indexes}
\label{sec:learnedindex}
Learned indexes build machine learning
models that predict the position of a given key.  
For example, one may train a linear model 
$pos=\floor{a \times key+b}$ where $a$ and $b$ are parameters learned from
prediction accuracy. A learned
index may use more complex models (e.g., neural networks), build a
hierarchy of simple models, or both.  Many learned
indexes~\cite{LearnedIndex,FITing-Tree,ALEX,RadixSpline,PGM,ZM,ML-Index,XIndex,LISA,RSMI,qd-tree,Tsunami,Flood,LIPP} are based on this idea, but most 
are read-only. We focus on updatable OLTP learned indexes, but to the best of our knowledge, 
none support persistence and very few support concurrency~\cite{XIndex}.

\textbf{Design Overview.}
We build APEX on ALEX~\cite{ALEX}, a fast, updatable learned index. 
We use ALEX to present our approach to crash consistency
and concurrency on PM. Many of our techniques are
applicable to other learned indexes; doing so is
interesting future work.
ALEX uses a hierarchy of simple linear regression models. 
The structure of ALEX is also called recursive model index (RMI).
It uses gapped arrays (i.e., arrays that leave empty slots between records to efficiently absorb future inserts) to store fixed-size keys and payloads sorted by keys.. 
As shown in Figure~\ref{fig:alex}, 
each inner node stores a linear model with $m$ child pointers ($m=4$ in Figure \ref{fig:alex}).  
Traversal starts from the root node which uses its linear model to predict the next child node (model) to probe, until reaching a leaf (data) node. 
Each data node stores a linear model and two aligned gapped arrays (GAs), one for keys and one for payloads to reduce search distance and cache misses (Figure~\ref{fig:alex} shows one for brevity).

\textbf{Models and Root-to-Leaf Traversals.}
ALEX uses ordinary least squares linear regression with closed-form formula to train data node models.  
We find that linear regression models work well in most datasets except one extremely non-linear data set (\fb in Section~\ref{sec:eval}). 
In such highly non-linear cases, complex models can possibly provide higher accuracy (and thus fewer PM accesses) but they also come with higher overhead for training and inference. 
How to balance model accuracy and the extra cost of complex models is an interesting future direction. 
Inner node models can partition key space flexibly. 
For example, in Figure~\ref{fig:alex} the root node model divides the key space $[0, 1)$ into four equally-sized subspaces, and each subspace is assigned to a child node; all keys in $[0, 1/4)$ are placed in data node A. 
In other words, inner node models do not \textit{predict} which child node a key falls in, but \textit{guide} how ALEX places keys in child nodes. 
Thus, model ``predictions'' in inner nodes during traversal are accurate by construction.

\textbf{Search.}
To probe a data node, ALEX uses the stored model to predict a position into the GA.
The search succeeds if the predicted position contains the target key. 
Otherwise, ALEX uses exponential search to find the key. 
If the keys are uniformly distributed (easy to fit by the model), and the number of keys is smaller than the maximum data node size, one may use a (large) data node to accommodate all records to drastically reduce inner node size. 
Otherwise, ALEX recursively partitions the key space to $m$ subspaces until the keys in each subspace can be modeled well by a linear model.
As Figure~\ref{fig:alex} shows, 
since subspace $[\frac{1}{4}, \frac{1}{2})$ is non-linear, another node is created hoping that the new subspaces are ``linear'' enough, 
while $[\frac{1}{2}, 1)$ is already linear, so ALEX uses one data node for it.

\textbf{Insert.}
An insert in data node first uses the model to predict the insert position in gapped array and may employ the exponential search to locate the proper position.
Two cases are possible upon insert: (1) insert into the dense region, or (2) insert to a gap.
Case (1) requires the elements shifts while case (2) needs to fill all consecutive gaps with the adjacent keys to enable exponential search; Both cases incur excessive PM writes.    
Such write amplification can easily saturate PM write bandwidth, limit the performance and make efficient crash consistency (without logging) impossible. 

\textbf{SMOs.}
ALEX defines node density as the fraction of filled GA slots, and further defines lower density $d_{l}$ (0.6 by default) and upper density $d_{u}$ (0.8 by default). 
Once the node's density is $> d_{u}$, an SMO is triggered, because insert performance will deteriorate with fewer gaps.
An SMO can expand or split a node. 
An expansion enlarges the node's GA.
So data nodes in ALEX are variable-sized. 
A split is carried out like in a \bptree. 
For example, when data node B in Figure~\ref{fig:alex} is split, two new nodes are allocated and trained with data partitioned across these two nodes.  
Then the two right-most pointers in the root node which originally point to B will respectively point to the two new data nodes. 
If node A is also split, there is no spare pointer in the root node. 
ALEX may double the root node's size or create a new inner node with two child data nodes (split downwards), each contains a split of A.
Deletion is simple because it can just leave a new gap.
ALEX may perform node contraction and merge to improve space utilization. 
ALEX uses built-in cost models to make SMO decisions using various statistics. 
More details about the cost model can be found elsewhere~\cite{ALEX}.  
Both inner and data nodes are variable-sized and can be much larger (e.g., up to 16MB) than nodes in a \bptree. 
Using large nodes is important for reducing tree depth, but may significantly slow down SMOs as model retraining takes more time and more data needs to be inserted to the new nodes. 
On multicore CPUs, this could present a scalability bottleneck as an SMO will block concurrent accesses to the node. 
Inner node SMO does not require model retraining. 
As noted earlier, inner node models are only used for space partition and are always accurate so that we only need to scale the model by doing simple multiplication.

\section{APEX Overview}
\label{sec:overview}
We design APEX with a set of principles distilled from the unique
properties of PM and learned indexes:

\begin{itemize}[leftmargin=*]\setlength\itemsep{0em}
\item \textbf{P1 - Avoid Excessive PM Reads and Writes.}  A
  practical PM index must scale well on multicore machines.  Given the
  limited and asymmetric bandwidth of PM, APEX must reduce
  unnecessary PM accesses and avoid write amplification.

\item \textbf{\textbf{P2} - Model-based Operations.}  Data-awareness
  and model-based operations uniquely make search operations
  efficient.  A persistent learned index such as APEX must retain this
  benefit.

\item \textbf{\textbf{P3} - Lightweight SMOs.}  Structural
  modification operations in learned indexes can be heavyweight and
  eventually 
  limit scalability.  APEX
  should be designed to reduce such overheads.

\item \textbf{\textbf{P4} - Judicious Use of DRAM.}  
APEX can use DRAM for performance, but should use it frugally to reduce cost.

\item \textbf{\textbf{P5} - Crash Consistency.}  APEX operations must
  be carefully designed to guarantee correct recovery.  Ideally, it
  should support instant recovery to achieve high availability.
\end{itemize}

\begin{figure}[t]
\centering
\includegraphics[width=0.93\columnwidth]{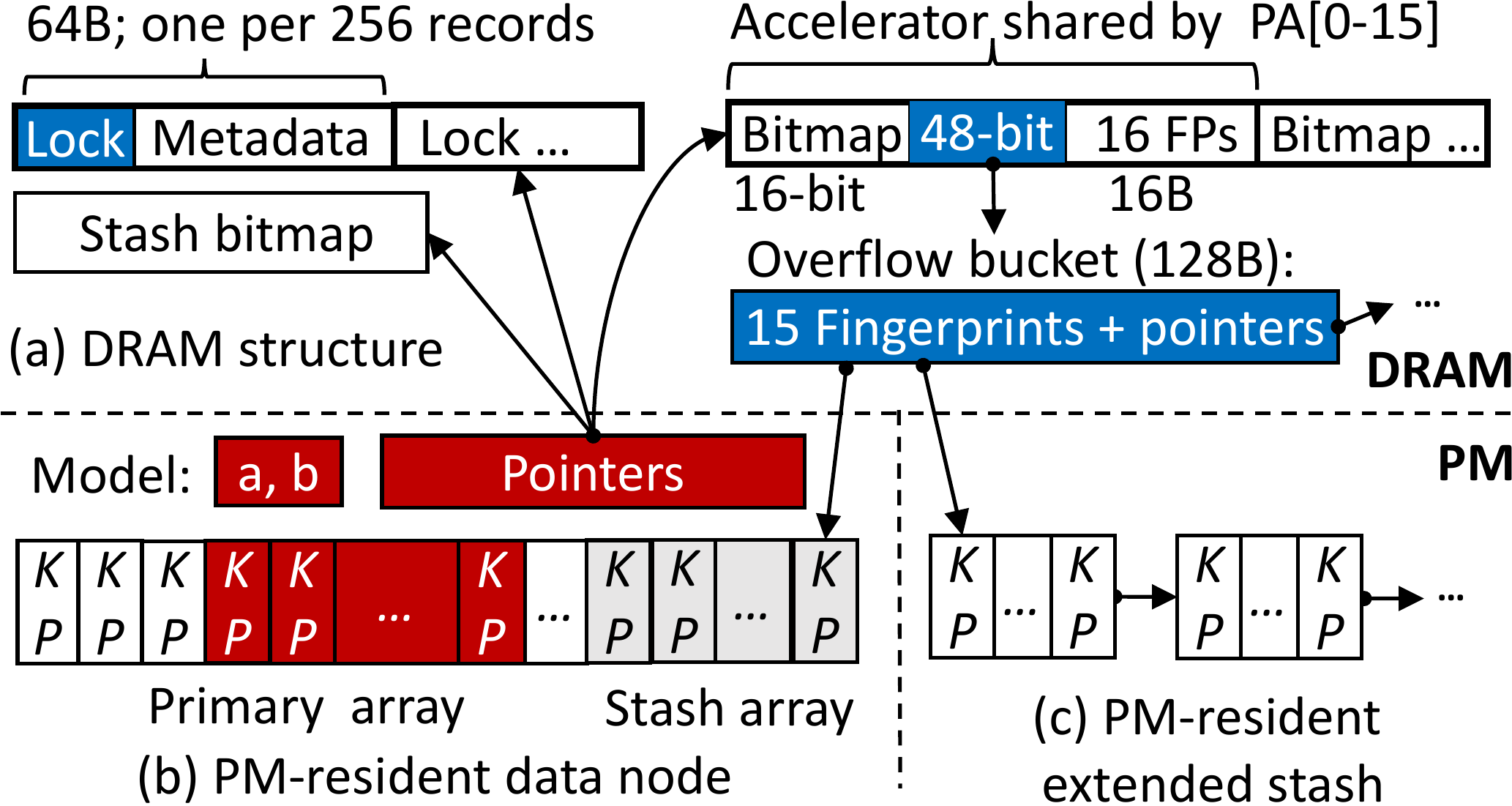}
\caption{APEX data node layout and DRAM-resident data. } 
\label{fig:tree} 
\end{figure}

\subsection{Design Highlights}
APEX combines new and existing techniques based
on the above design principles. Similar to ALEX~\cite{ALEX}, APEX consists of inner nodes and data nodes. 
APEX places all node contents in PM except
a small amount of metadata and accelerators in DRAM to improve performance and reduce PM writes (\textbf{P1}, \textbf{P4}).  
APEX employs model-based insert~\cite{ALEX} where each data node can be treated as 
a hash table that uses a model as an order-preserving hash
function to predict insert location. 
To resolve collisions without introducing unnecessary PM accesses, we propose a
new probe-and-stash mechanism (Section~\ref{sec:ins}) inspired by recent
PM hash tables~\cite{Dash} (\textbf{P1}, \textbf{P2}). 
We set different maximum node sizes for APEX's inner and data nodes to ensure most SMOs do not 
hinder scalability while maintaining a shallow tree (\textbf{P3}).  
For instant recovery, we design DRAM-resident
components to be reconstructable on-demand (\textbf{P5}).  

\subsection{Node Structure}    
\label{sec:datanode}
Each node in APEX contains a linear model consisting of two double-precision floating point values (slope and intercept) stored in node header, e.g., \texttt{a}, \texttt{b} in the data node in Figure~\ref{fig:tree}(b). 
Each inner node also contains an array of child pointers.  
Data nodes also store key-payload pairs as records. 
Same as other learned indexes~\cite{ALEX,PGM,XIndex,RadixSpline},  
APEX stores fixed-length\footnote{There has been initial work supporting variable-length keys~\cite{sindex,umar21}. 
As future work, we hope to explore how APEX could adopt these techniques.}
numeric keys that are at
most 8-bytes and 8-byte payloads (either inlined or a pointer). 
Like previous work~\cite{LIPP,Dash}, we assume unique keys, but non-unique keys can be supported by 
storing a pointer to a linked list of records as payload. 

Data nodes in APEX are variable-sized, but have a fixed \textit{maximum} size which is set to 256KB to fully exploit models. 
This is larger than the typical size (256 bytes -- 1KB) in B+-trees, but small enough to efficiently implement SMOs and achieve good scalability. 
Since SMOs in inner nodes are relatively rare and 
exhibit low SMO cost (Section~\ref{sec:eval}), we keep the maximum size of inner nodes to be 16MB. 
This gives APEX more flexibility to select node fanout, lower tree depth and maintain good search performance. 
Because of the low tree depth, inner nodes exhibit good CPU cache utilization. 
Placing them in DRAM does not benefit much. 
Thus, different from PM-DRAM B+-trees~\cite{FPTree,lbtree}, we place inner nodes in PM, which also enables instant recovery (Section~\ref{sec:cc}).

To support model-based lookups and hash-based inserts, each data node
consists of (1) a primary array and a stash array (and in case
of overflows, extended stash blocks), which store records and are
PM-resident, and (2) reconstructable metadata stored in DRAM to
accelerate various operations and to support concurrency.  

\textbf{PM-Resident Primary and Stash Arrays.}  
As shown in Figure~\ref{fig:tree}(b), both arrays store data records in record-sized slots. The linear model predicts
a position in the primary array (PA) for a given key. To insert a
new record $\langle K, P \rangle$, if the predicted position in the
PA is not free, APEX linearly probes the PA 
and inserts the record into the first free slot. 
We limit the probing distance to a constant $D=16$ so that the thread would
probe no more than two XPLines (512 bytes).
Bounding the probing distance also allows records in PA to be
nearly-sorted~\cite{nearsort}, improving search performance. 
If no free slot is found, APEX inserts the new record to a free slot in the
stash array (SA), which acts as an overflow area.  
Stashing allows APEX to efficiently resolve collisions without excessive PM writes compared to using ALEX's gapped array (element shifts) or other common techniques 
(e.g., probing a large number of slots with excessive PM accesses). 
We also use accelerators to reduce the overhead of stash accesses.
We present more details in Sections~\ref{sec:search} and \ref{sec:ins}. 

Each node has a determined size and number of record slots (computed using the number of keys in the node and lower density),
so APEX needs to properly divide the slots between the PA and SA. 
Allocating more slots to PA can lower collision rate (faster
search and inserts), yet there must be enough stash slots in case
collisions do happen. Therefore, APEX needs to strike a balance between
insert/probe performance and collision handling. APEX leverages data
distribution to solve this problem 
(Sections~\ref{sec:smo_stash_ratio} and~\ref{sec:bulk_load}).

\textbf{DRAM-Resident Metadata and Accelerators.} 
APEX places in DRAM certain structures that are (1) easy to reconstruct in
case of failures yet are (2) very critical to performance at runtime. 
As Figure~\ref{fig:tree}(a) shows, we store metadata, locks and accelerators 
in DRAM, accessible via the pointers stored in the PM data node's header.
Metadata includes basic information about the node, e.g., number of records. 
As we discuss later, locks do not need to survive power cycles for recovery,
and placing them in DRAM can avoid excessive PM accesses. 
Accelerators are compact data structures to enable fast record access with reduced PM accesses. 
The key is to use fingerprints~\cite{FPTree} to quickly determine if a key possibly exists (often without even
reading the whole key). 
Bitmap indicating slot status is also used for inserts to quickly locate a free PA slot. 
Finally, to reduce storage overhead we share an
accelerator for every 16 records in the PA (\texttt{pa}), e.g., in Figure
\ref{fig:tree}, \texttt{pa[0]}--\texttt{pa[15]} share the first
accelerator. 
One accelerator (24-byte) includes 16-byte fingerprints, a 16-bit free-slot bitmap, and one 
48-bit pointer\footnote{Modern x86 processors use the least significant 48 bits for addressing~\cite{IntelManual}}. 
A stash bitmap is used to indicate the empty slots in the stash array.

\section{APEX Operations}
\label{sec:operation}
Now we present APEX operations in a single-thread setting with crash-consistency. 
Section~\ref{sec:cc} discusses concurrency and recovery.

\subsection{Search \& Range Query}
\label{sec:search}
To search for a key, we start at the
root node and use its model to predict which child node to go
to, until we reach a leaf node. 
There is no search within inner nodes as by construction
the model's prediction is always accurate. 
All operations that require a traversal share this logic; for now we focus on data node operations.
Within a given data node, we devise a probe-and-stash mechanism to
reduce unnecessary PM accesses during key lookups. 
As Figure~\ref{fig:tree} shows, all memory accesses by a lookup are highlighted in red (blue) for PM (DRAM). 
We first ``probe'' using the node's model to
predict the position of a key $k$ in the PA. Suppose
the predicted position is 3 in Figure~\ref{fig:tree}(b). 
APEX directly returns the record if the key is found in the slot. 
Otherwise, it linearly probes from $PA[3]$ till
$PA[3+D-1]=PA[18]$ where $D$ is the probing distance (16) described in
Section~\ref{sec:datanode}. 
APEX returns the record if $k$ is found;
otherwise it continues to check the SA. 
Note that for all operations linear probing always proceeds in the same direction (conceptually from ``left to right'')
because APEX does not shift data during inserts and so a key cannot be stored in a slot before its predicted position, simplifying concurrency control (details later). 

To accelerate the lookup in stashes, APEX creates an overflow
bucket in DRAM for every 16 PA records if a key is
originally predicted within the 16 records but overflowed to SA or extended stash; 
the overflow bucket's address is stored in the DRAM 
accelerator. 
We find the value of 16 
balances memory consumption and performance: having an overflow
bucket per record needs a 48-bit pointer (next to
the bitmap in Figure~\ref{fig:tree}(a) in the accelerator) to
point to it, adding non-trivial overhead ($>$
37\%). Using one overflow bucket per 16 records amortizes this cost. 
Continuing with the example, before accessing the stash, APEX first checks
the corresponding overflow bucket, which  
holds up to 15 pointers for indexing overflow keys in the stash.  
A new overflow bucket is dynamically allocated and linked if there are more than 15 overflow keys.

Each pointer to a stash record in the overflow bucket inlines a 
fingerprint of that stash record in the most significant byte of the pointer.  
Only keys in the stash with matching fingerprints 
are accessed from PM.  
A negative search 
will issue no PM reads on the stash, but up to two
XPLine accesses in the PA. For a positive search,
fingerprints in the overflow
buckets pinpoint the target stash entries, reducing
the expected number of PM reads to one ~\cite{FPTree}.

For a range query where $i$/$j$ are the
predicted positions of the start/end keys in a data node, APEX first collects the
records between $PA[i]$ and $PA[j+15]$, and 
finds the remaining ones from stash. 
Indexes that use unsorted
nodes~\cite{lbtree,BzTree}, often require sorting full nodes,
which adds overhead and requires using smaller 
nodes to alleviate. APEX only needs to sort the final result set. 
This is efficient as APEX maintains
nearly-sorted order, sorting which is
faster~\cite{nearsort}. In Section~\ref{sec:eval} we show its impact
using realistic datasets.

\subsection{Insert}
\label{sec:ins}
To insert a record with key $K$, APEX ensures 
$K$ is not already in the index, and if so, locates and inserts the record to a free slot.

\textbf{Uniqueness Check.}
After obtaining a predicted position $k$, instead of directly probing
PA, we first check key existence using the
fingerprints in the accelerator. This can potentially
save PM accesses: a negative result indicates $K$
is definitely not in PA. Fingerprints 
are usually co-located in the same cacheline with the 16-bit bitmap which  
must be brought to the CPU
cache to find a free slot for PA insertion, accessing
fingerprints incurs little overhead and is practically ``free''
without additional memory accesses.
In Figure~\ref{fig:tree}, assume the model
predicted position 3 for the new key,  
APEX first checks 
existence using the 4th to 16th fingerprints from the first
accelerator, and the first three fingerprints from the second
accelerator. We access PA only if there is a matching
fingerprint.
Note that lookup (Section~\ref{sec:search}) does not use PA's fingerprint because first accessing the accelerator can incur extra cache misses if the key is stored in PA (which is the common case). 
The uniqueness check then continues with the fingerprints in the
overflow bucket(s) and (if needed) stash slots in PM, same as a
regular key search. 

{\bf{Locating a Free Slot.}} 
PM indexes often use bitmaps to indicate free space~\cite{FPTree,lbtree,dptree,Chen2015}, but persisting them for each insert/delete on PM 
adds non-trivial overhead. 
Thus, APEX includes the bitmaps in 
the DRAM accelerators that are rebuilt on-demand by reading slot contents. 
We indicate free slots by storing in them an invalid key that is out of the node's key range $[min, max]$.
Then we must not place {$min~Int64$} and $max~Int64$ in the same node. 
This is done by ensuring the initialization/bulk loading algorithm always generates at least two data nodes, one 
with range $[min~Int64, max]$, and the other with range $[min, max~Int64]$.
Then we use $max + 1$ and $min - 1$ as the invalid key in the two nodes, respectively.
As a result, both insert and delete only require one PM write
(updating the record) and one DRAM write (flipping a bitmap entry). 

SA is statically allocated during node creation. 
When it is full (although rare), APEX dynamically allocates a new
256-byte extended stash in PM and atomically stores a pointer to
the newly allocated block in PM. 
Extended stash blocks are linked together and reachable via a pointer in the data node. 
Crash consistency is guaranteed by the PM allocator~\cite{PMDK} 
which ensures safe PM ownership transfer between the allocator and
PM to avoid permanently leaking PM. 

{\bf{Crash-Consistent Insert.}} 
As described in Section~\ref{sec:bg}, ALEX~\cite{ALEX} uses GA with exponential search, which incurs excessive PM writes by shifting records or filling consecutive gaps with adjacent keys. 
To save PM bandwidth, APEX neither shifts records nor fills gaps. 
This is possible since APEX uses linear probing instead of exponential search with two careful designs:
(1) APEX co-locates the key and payload, so an insert requires only a 16-byte PM write. 
(2) APEX writes the payload \textit{prior to} writing the key and persists them in PM using one flush and fence, leveraging the fact that modern x86 CPUs do not reorder writes to the same cacheline~\cite{FlogWrite}. 
In case of a crash, APEX simply discards records with invalid keys. 
Thus, unless a new extended stash block is needed, an insertion only writes one XPLine.
We alleviate the impact of extended stash by carefully setting SA size based on
data distribution (Sections~\ref{sec:smo_stash_ratio}--\ref{sec:bulk_load}). 

\subsection{Delete \& Update}
\label{sec:delupd}
APEX implements delete as lookup followed by invalidation.
Once we locate the record, APEX simply replaces the target key in the slot with an invalid key. 
The validity bitmap in DRAM is also updated to reflect this change. 
APEX updates records in place. If the key is found, we atomically update and flush the payload with one XPLine write.
Same as inserts, the lookup process in deletes and updates uses PA's fingerprints to reduce PM accesses.

\subsection{Structural Modification Operations}  
\label{sec:smo}
Similar to ALEX, APEX uses node density to decide when to
trigger an SMO.  ALEX uses 0.8 as the upper density limit $d_u$ because
insert performance degrades beyond that. To achieve an average memory
utilization of $\sim$70\% (same as a \bptree), ALEX uses 0.6 as the
lower density limit $d_l$.
In APEX, however, such a tight bound would trigger many SMOs, incur excessive PM
writes (e.g,. moving data to a new node for node splits) and hurt
performance.  Since inserts in APEX incur little write amplification, APEX
can tolerate a higher upper density limit to reduce SMOs. 
Based on empirical evaluation, we use $[0.5, 0.9]$ with the same $\sim$70\%
average memory utilization.

\textbf{Node Expansion vs. Split.}
Once an SMO is triggered, we use cost models to choose between
node \emph{expansion} and \emph{split} like ALEX does. 
APEX follows the same model as ALEX's but uses different
statistics. 
To quantify the cost of a search, APEX uses the average number of cache misses in 
probe-and-stash instead of ALEX's average number of iterations of
exponential search.  Insert cost is
the average number of overflow buckets allocated plus the search cost.

\textbf{Data Node Expansion.} 
APEX expands a node $A$ in three steps: (1) allocate and
initialize a new node $B$; (2) retrain or re-scale the
model and insert records from $A$ to $B$ using the new
model; (3) attach $B$ to the parent node, update the sibling pointers,
and reclaim $A$.
 
This multi-step process needs to be implemented carefully.  For
example, there will be a PM leak if a crash happens before step~3.
Also, the index would be inconsistent if a crash happens during
step 3.  APEX achieves lightweight crash-consistency 
via \emph{hybrid logging}.  We make a key
observation: only step~2 is relatively long running.  Hence, different
from prior work which only uses redo-logging to possibly redo a 
\emph{whole} SMO, APEX uses undo logging before step 3 and
redo-logging after step 2.  If a crash happens after 
step 2, APEX would waste less work and can resume step 3
upon restart.

A na\"ive logging approach, such as PMDK's physical undo logging logs
all data records and so incurs excessive PM writes. In APEX, 
node expansions have only three steps, so we can use logical logging with
a small log area in PM. Upon step 1, we
initialize a ``node-expand'' log entry (one cacheline) in PM with the format of 
$\langle old, new, key,
stage\rangle$, where $old$ is a pointer to the old node, $key$ is the insert
that triggered the node expansion (for locating the parent
node of the old node), and $stage$ is set to \texttt{UNDO}. Next, we allocate a
new node using PMDK which atomically stores the new node's address in $new$, 
and 
initialize the new node (setting all keys in the
PA/SA to invalid). 
Step 2 then updates and persists the $stage$ of
the node-expand log entry to \texttt{REDO}. 
Now we switch from undo logging to redo logging and start step 3.
Finally, we persists the node-expand log entry in the PM
with a $stage$ value reset to \texttt{NoSMO}.
 
This approach gives low overhead (only three PM log writes) and 
fast recovery.  If the system fails before step 3, we discard 
the new node to undo the incomplete expansion. 
If it fails after step 2, the SMO can resume from step
3 by observing $stage$ in the log.

\textbf{Data Node Split and Inner Node Expansion.}
APEX uses the same data node split and inner node expansion logic
as ALEX (Section~\ref{sec:bg}). APEX also handles these SMOs in a very similar way to node
expansions explained above: Each SMO has its own log entry in PM.
We use logical undo-redo logging to ensure once a heavyweight step
(e.g,. record copying) is done, APEX would only redo the lightweight
step (e.g., switching pointers) upon recovery.

 \textbf{Stash Ratio.}
 \label{sec:smo_stash_ratio}
The ratio
between the sizes of the stash and primary array is governed by a \emph{stash ratio} $S$, defined as the fraction of stash array size to the sum of the primary array and stash array sizes. 
 Setting a reasonable stash ratio is needed to avoid
excessive collisions or the overhead of extra stash block
allocations. 
Previous PM hash tables~\cite{Dash,LevelHashing} allocate a
fixed-size stash array based on a predefined collision
probability. 
This is not ideal for learned indexes since the collision probability of the model depends on how well the model fits the data. 
APEX automatically configures the stash ratio when
creating a new data node, based on the overflow ratio of the old node.
Specifically, the overflow ratio $O$ of a data node with $N_d$
keys is the fraction of the number of overflowed keys $N_o$ in that
data node, i.e., $O = \frac{N_o}{N_d}$.  These simple statistics are
all part of the metadata in DRAM and APEX maintains a set of them 
per 256 records to amortize the cost. APEX assumes that the expanded or split nodes
follow the same distribution from the old node.  Hence, the stash
ratio of new node created by an SMO is set to be the overflow ratio.
This strategy ensures that the stash ratio of a data node is adaptive
to the actual data distribution.

\subsection{Bulk Loading}
\label{sec:bulk_load}
Like ALEX, during bulk loading APEX grows the RMI greedily, 
but uses different cost models described in Section~\ref{sec:smo} 
and must also determine PA and SA sizes. 
Ideally the stash ratio $S$ should match the percentage of records overflowed to SA during real inserts to reduce extended stash use and balance insert (which prefers larger SA) and lookup (which prefers smaller SA) speeds. 
This requires knowing data distribution which is unavailable upon bulk loading. 

We estimate a reasonable $S$ empirically. 
Based on extensive experiments using realistic datasets (details in Section~\ref{sec:eval}), we find setting $S$ within the range [0.05, 0.3] well balances model-based search and insert performance. 
APEX thus bounds $S$ in this range. 
Then, we set $S$ via simulation: 
Given $N_d$ keys to insert to a node, we first assume the node will reach the upper density limit $d_u$ and all free slots are allocated to PA. 
We then compute the predicted positions in PA for each key and probe-and-stash to collect the number of keys $N_o$ overflowed to SA without actually carrying out any inserts (thus a ``simulation'').  
With the overflow ratio $O=\frac{N_o}{N_d}$, 
we calculate $S$ based on two intuitions:
(1) the higher the overflow ratio $O$, the bigger the stash ratio $S$;
(2) $S$ should be greater than $O$ as $O$ was determined by assuming all slots are in PA (i.e., real inserts should exhibit more collisions than simulation). 
There could be many ways to determine $S$ using $O$. 
For simplicity, we set $S$ to be a multiple ($n$) of $O$, i.e., $S=n\times O$, and empirically determined $n$'s value to be 1.5 via experiments (not shown here for space limitation); we call $n$ the ``stash coefficient.'' 
In general, a higher stash coefficient means potentially more keys are stored in the stash areas. 
Taking the aforementioned bound into account, $S = max(0.05, min(0.3, 1.5\times O))$.

Overall, our method gives reasonable performance and is simple to implement/calculate. 
The upper limit ensures most records stay in PA; 
the lower limit is a safety net to absorb collisions (e.g., due to a distribution shift) before an SMO reorganizes the node. 
In practice, we do not expect stashing to be the main storage as APEX recursively partitions the key space so that each subspace can be modeled well. 
Note that bulk loading still succeeds even if $S$ is inaccurate: more keys will be stored in SA and/or the extended stash.
Our evaluation in Section~\ref{sec:eval} shows that in practice stash ratio is low in most cases and extended stash blocks are rarely used as stash ratio is reasonably set based on data distribution.
Further optimizations are interesting future work.

\section{Concurrency and Recovery} 
\label{sec:cc}
As Section~\ref{subsec:pmtree} describes, compared to traditional node-level locking and lock-free approaches, optimistic locking is usually a better fit for PM and balances programmability and performance.  
APEX further adapts optimistic locking for learned indexes on PM. 

\textbf{Inner Node Accesses.}
APEX uses different maximum sizes for inner and data nodes (16MB vs. 256KB). 
Inner nodes typically have less contention so we pick a larger node size. 
Each inner node carries a reader-writer lock for SMO, compared to traditional optimistic locking with mutual exclusion locks~\cite{OLC}. 
Reading an inner node (e.g., lookup) is lock-free. 
In traditional optimistic locking, the thread retries traversal if inconsistencies caused by modifications on the node are detected, wasting CPU cycles.
APEX avoids such aborts by an out-of-place-based SMO design, described later. 

\textbf{Data Node Accesses.}
Data nodes may see many concurrent accesses, so using a smaller node size can help reduce contention. 
In addition to a node-level lock to ensure only one thread can conduct SMO on the node, we allocate one optimistic lock per 256 records in PA to isolate non-SMO updates.
This design balances the synchronization and lock acquisition overhead during SMOs. 

To read a data node, the thread keeps traversing down until reaching the target data node, without holding any locks. 
However, upon reading data records, it uses the version in the optimistic lock to guarantee the read correctness and restarts the search if the version changed. 
Like many prior approaches, we use epoch-based memory reclamation~\cite{epoch} for safe memory management. 

To insert a key, the thread first traverses to the target data node using lock-free read. 
To find the key in the node, the thread may need to use linear probing to access multiple slots, which requires acquiring the corresponding lock(s) that cover(s) the probing slots. 
More than one lock may be acquired if the predicted position plus probing distance crosses lock boundary.  
Unlike ALEX, since in APEX all threads linearly probe in the same direction (described in Section~\ref{sec:search}), it is guaranteed that deadlocks will not happen.  
To update a data node, the thread first acquires the lock that protects the record, and then continues to hold it if it needs to update the stash array. 
Multiple threads can race to install new records into the stash array while holding different locks (i.e., in two different 256-record blocks). 
Therefore, threads must first allocate a free stash slot using the stash bitmap, which is done by using the compare-and-swap (CAS) instructions to atomically set the ``next free'' bit in the bitmap (and retry if the CAS failed); 
after that each thread can continue to insert the record to its own stash array slot.  

\textbf{SMOs.}
Data node expansion and split require more care to work under optimistic locking. 
Expanding a data node is done in an out-of-place manner that always allocates a new node and updates the parent node to point to the new node. 
Meanwhile, the parent node may be undergoing an expansion. 
The aforementioned reader-writer lock in inner nodes is for handling such cases. 
Upon updating the inner node, the thread $T$ takes the node's lock in shared (reader) mode. 
Note that since $T$ already holds the lock for the data node, it is safe to directly use an atomic write to update the pointer in the parent node. 
This allows multiple threads to proceed and expand different data nodes in parallel. 
If a data node split causes the parent node to expand, the inserter thread locks the inner node in exclusive (writer) mode. 
In theory, it is possible for splits to propagate up and grow the RMI by acquiring locks bottom-up. 
This can incur non-trivial overhead~\cite{ALEX}. 
Our implementation therefore follows prior work~\cite{ALEX} to disallow inner node split and only allow expansion. 
This limits the number of acquired locks to three (data node, parent and grandparent levels). 
Note that throughout this process, readers proceed without taking any locks but must verify version numbers. 
Using out-of-place SMO and updates without shifting in inner nodes allow the traversals to data nodes without retries. 
The thread may see an obsolete data node due to concurrent SMOs (although the key range is correct).  
It detects this case by checking data nodes's lock status and retries from root if it is set.

\textbf{Instant Recovery.} 
APEX adopts lazy recovery in Section~\ref{subsec:pmtree}. 
It needs to undo in-flight SMOs (if any) by deallocating PM blocks and switching pointers (Section~\ref{sec:smo}); both are lightweight and after that APEX can start to handle requests. 
Since each thread has at most one in-flight SMO upon crash, recovery time scales with thread count, instead of data size. 
Modern OLTP systems usually limit thread count, making APEX recovery practically instant.

\section{Evaluation}
\label{sec:eval}
We now present a comprehensive evaluation of APEX including
comparisons against the state-of-the-art PM indexes. We show that:
\begin{itemize}[leftmargin=*]\setlength\itemsep{0em}
\item APEX retains the benefits of model-based search and achieves high throughput and good scalability. 
\item APEX's individual design principles and choices are effective, collectively allowing APEX to perform and scale well.
\item APEX instant recovers ($<$1s), although it uses DRAM, in contrast to prior work that trades off instant recovery for performance. 
\end{itemize}

\begin{figure*}[t]
\centering
\includegraphics[width=\textwidth]{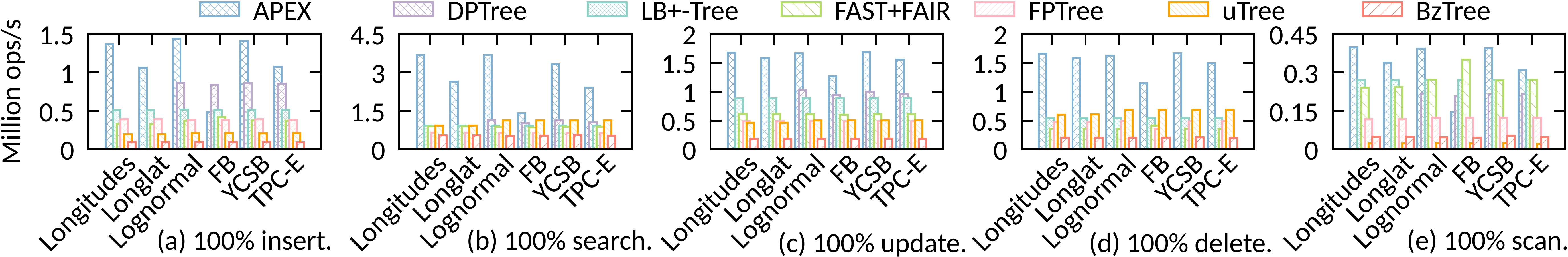}
\caption{Single-thread throughput. 
APEX performs the best on five datasets for inserts and range scans, and remains competitive for the worst-case \fb.
For search, update and delete APEX performs the best across all cases.}
\label{fig:single-perf}
\end{figure*}

\subsection{Index Implementations} 
We implemented APEX in C++ 
and compare it with recent PM B+-trees: BzTree~\cite{BzTree}, \lbtree~\cite{lbtree}, FAST+FAIR~\cite{Hwang2018}, DPTree~\cite{dptree}, FPTree~\cite{FPTree} and uTree~\cite{utree}. 
BzTree and FAST+FAIR are PM-only indexes and do not use DRAM.  
\lbtree and \fptree are hybrid PM-DRAM indexes that place inner/leaf nodes in DRAM/PM. 
They combine HTM and locking for synchronization. 
\utree puts both inner and leaf nodes in DRAM and relies on a linked list of records in PM for persistence. 
\dptree batches modifications in DRAM buffer and merges with the background PM-DRAM tree to reduce persistence overhead.
Except BzTree and FPTree (which were not originally open-sourced), 
we use the original authors' open-sourced code and add the necessary but missing functionality in best effort.\footnote{Code for all indexes is summarized in our repo: \url{https://github.com/baotonglu/apex}.}

\textbf{Persistence.} 
We use PMDK~\cite{PMDK} to support persistence in all indexes and verify they do not incur unnecessary overheads. 
We modified \lbtree, FAST+FAIR, uTree and DPTree to use PMDK because they either were proposed based on DRAM emulation or did not implement certain necessary PM-related functionality due to the lack of a full-fledged PM allocator.\footnote{For example, \utree was designed to self-manage PM space, but the open-sourced code does not implement recycling. So we use the PMDK allocator.}
We also fixed \lbtree to provide correct read committed isolation level like other indexes.\footnote{Details at \url{https://github.com/schencoding/lbtree/pull/6}.}

\textbf{Operations.} We did best-effort implementations of the
missing range scan in FAST+FAIR, \lbtree and \utree.  
We faithfully implemented recovery for \lbtree and \utree.  
For multi-threaded recovery, \lbtree requires statistics~\cite{lbtree} that are currently not being collected.  
We therefore implemented a single-threaded version.

\subsection{Experimental Setup}
\label{sec:setup}
We run experiments on a server with a 24-core (48-hyperthread), 2.1GHz Intel Xeon Gold 6252
CPU, 768GB Optane DCPMM ($6\times128$GB DIMMs on
all six channels) and 192GB DRAM ($6\times32$GB DIMMs). The CPU has
35.75MB of L3 cache.  
The server runs Arch Linux (kernel 5.10.11). 
We use PMDK/jemalloc~\cite{jemalloc} to allocate PM/DRAM. 
All code is compiled with GCC 10.2 with all optimizations. 
For fair comparison, we set each index to 
use the parameters used in its original paper.  
\lbtree/FAST+FAIR/\utree/BzTree use 256B/512B/512B/1KB node. 
\fptree uses 28/64-record inner/leaf nodes.
APEX uses maximum 16MB/256KB inner/data nodes. 

\textbf{Datasets.}
We use six synthetic and realistic datasets to test all the indexes. 
\longitudes is extracted from Open Street Maps
(OSM)~\cite{OSM}. 
\longlat is also from OSM but is transformed to become highly non-linear to stress learned indexes.  
\lognormal represents the lognormal distribution. 
\ycsb contains user IDs in YCSB~\cite{YCSB}.
SOSD~\cite{SOSD} includes four realistic datasets. Due to space
limits, we focus on the Facebook (\fb) dataset containing
randomly sampled Facebook user IDs.  \fb is extremely non-linear and
the hardest-to-fit among SOSD datasets. We use it to stress test the indexes. 
We also run the TPC-E~\cite{TPCE} benchmark and collect three datasets (trade, settlement, cash transaction) by loading the database with 15000 customers and 300 initial trading days. 
APEX performs similarly under them, so we only report results from the trade dataset. 

All keys are unique in these datasets. 
Same as previous work~\cite{ALEX} we randomly shuffle them to simulate a uniform distribution over time.
All the datasets use 8-byte keys and 8-byte payloads. 
Except \longitudes and \longlat whose key type is \texttt{double}, all the other datasets consist of 8-byte integer keys. 
\lognormal contains 190 million keys (2.83GB), whereas \tpce (trade) contains 259 million keys (3.86GB); other datasets include 200 million keys (2.98GB). 

\textbf{Benchmarks.}  We stress test each index using microbenchmarks.
For all runs, we bulk load the index with 100 million records, and then 
test individual
operations. Since only
\lbtree supports node merge (when the node is empty) which may
reduce its performance, we run 90 million deletes to avoid
triggering merges for fair comparison. Other 
workloads issue 100 million requests.  Range scans start at a random
key and scan 100 records.  \lognormal only has 190 million
keys, so for its insert test we issue 90 million requests.
The source code of DPTree does not support \texttt{double} key type (used in \longlat and \longitudes datasets), recovery and delete operation, so we do not include it in the corresponding experiments.

\subsection{Single-thread Performance}
\label{sec:single-thread}
We first run single-thread experiments to compare 
the indexes without contention.  Note that we do not remove
concurrency support; with one thread the overhead is 
minimal.  
To better understand the results, we also show the basic statistics of APEX after bulk loading in Table~\ref{tab:bulk-statistic}.
For inserts, as shown in Figure~\ref{fig:single-perf}(a), 
APEX outperforms
BzTree/\lbtree/FAST+FAIR/\dptree/\utree/\fptree by up to 15$\times$/2.7$\times$/3.8$\times$/1.66$\times$/6.8$\times$/3.7$\times$,
for five datasets.  The advantage mainly
comes from APEX's model-based insert and probe-and-stash.  In
most cases, APEX only issues one PM write per insert,
whereas other indexes often issue more (e.g., $>10$ for BzTree~\cite{BzTree}).  
For \fb, APEX is 
1.18$\times$/5.16$\times$/2.3$\times$/1.26$\times$ faster than FAST+FAIR/BzTree/\utree/\fptree, but achieves a lower (5.77\%/42\%) throughput when compared to \lbtree/\dptree.
As Table~\ref{tab:bulk-statistic} shows, \fb exhibits the highest average overflow ratio.  
Thus, \fb incurs many collisions in PA with more inserts routed to SA and extended stash. 
This requires more CPU cycles to find a free slot and allocate 
overflow buckets. Note that \fb is the hardest-to-fit (the worst case). 
Overall, APEX remains competitive under worst-case scenarios and outperforms 
other indexes by up to 15$\times$ in common cases.

\begin{table}[t]
\scriptsize
\centering
\caption{APEX statistics after bulk loading.} 
\smallskip\noindent
\setlength\tabcolsep{1.2mm}
\begin{tabular}{lrrrrrrr}
\toprule 
\textbf{Metric} & \textbf{\longitudes} & \textbf{\longlat} & \textbf{\lognormal} & \textbf{\fb} & \textbf{\ycsb} & \textbf{\tpce} \\
\midrule  
Average depth& 1.10& 1.64& 1.95& 3.13& 2 & 3.43\\
Maximum depth& 2 & 3 & 3& 6& 2 & 9 \\
Number of inner nodes& 541 & 3628 & 374& 6279 & 8193 & 20879 \\
Number of data nodes& 17438 & 44071& 12696 & 81856 & 16384 & 143035 \\
Minimum inner node size& 16B & 16B & 16B & 16B & 16B & 16B \\
Median inner node size& 16B & 32B & 64B& 32B& 16B & 16B \\
Maximum inner node size& 512KB & 4MB & 16MB & 512KB& 64KB & 4KB \\
Minimum data node size& 496B & 496B & 496B& 496B & 131KB & 1056B \\
Median data node size& 139.5KB & 29.05KB & 168.75KB& 21.5KB& 136KB & 2.56KB \\
Maximum data node size& 256KB & 256KB & 256KB & 256KB& 142KB & 149KB \\
Average stash ratio& 0.05& 0.118 & 0.05 & 0.299& 0.05 & 0.06 \\
Average overflow ratio& 0.006& 0.065 & 0.0008 & 0.48& 0.0014& 0.034 \\
\bottomrule 
\end{tabular}

\label{tab:bulk-statistic}
\end{table}

\begin{figure*}[t]
\centering
\includegraphics[width=\textwidth]{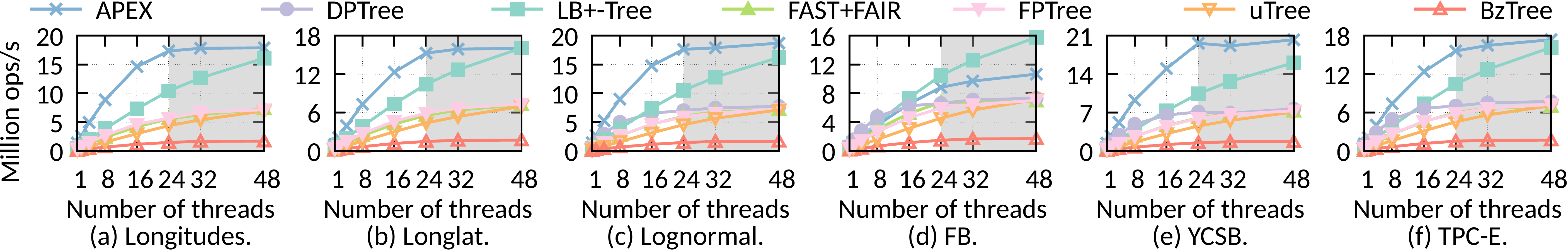}
\caption{Insert scalability. 
APEX and \lbtree scale better than others over all datasets. 
\lbtree performs better than APEX on \fb, but is limited by its lower single-thread throughput for other datasets.}
\label{fig:insert-scale}
\end{figure*}

\begin{figure*}[t]
\centering
\includegraphics[width=\textwidth]{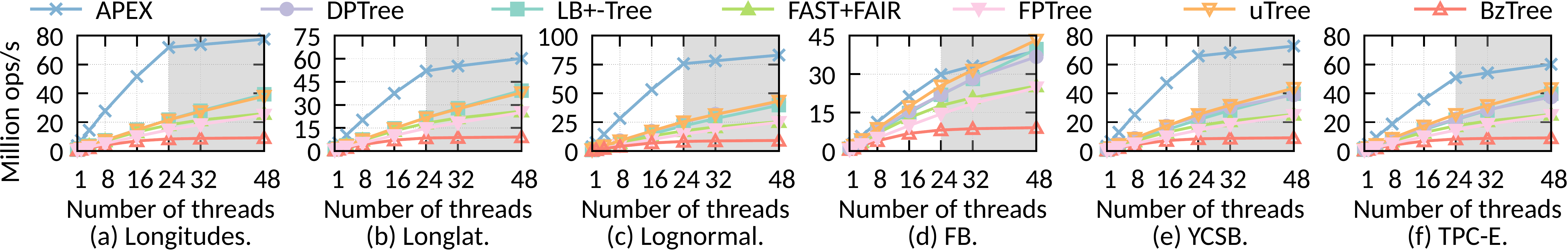}
\caption{Search scalability. 
APEX scales nearly linearly before hyperthreading, as model-based search leverages CPU caches more effectively than data-agnostic designs where tree traversal incurs many more cache misses.}
\label{fig:search-scale}
\end{figure*}

\begin{figure*}[t]
\centering
\includegraphics[width=\textwidth]{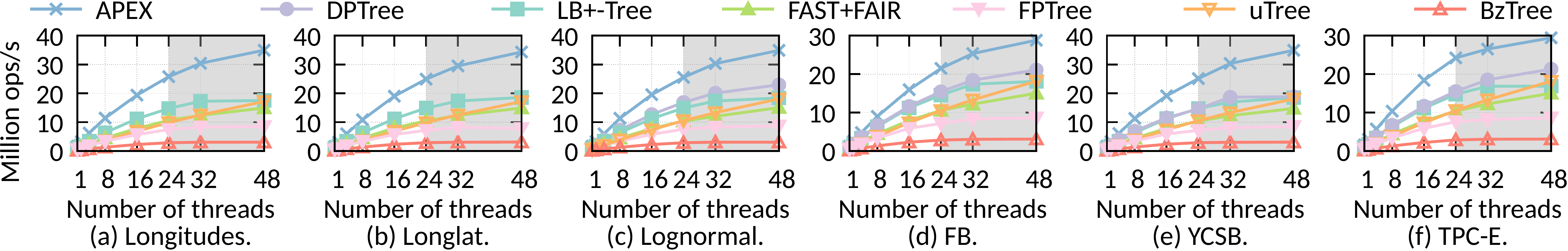}
\caption{Update scalability.
APEX issues one PM write per update, and so reduces PM bandwidth consumption and scales better. }
\label{fig:update-scale}
\end{figure*}

\begin{figure*}[t]
\centering
\includegraphics[width=\textwidth]{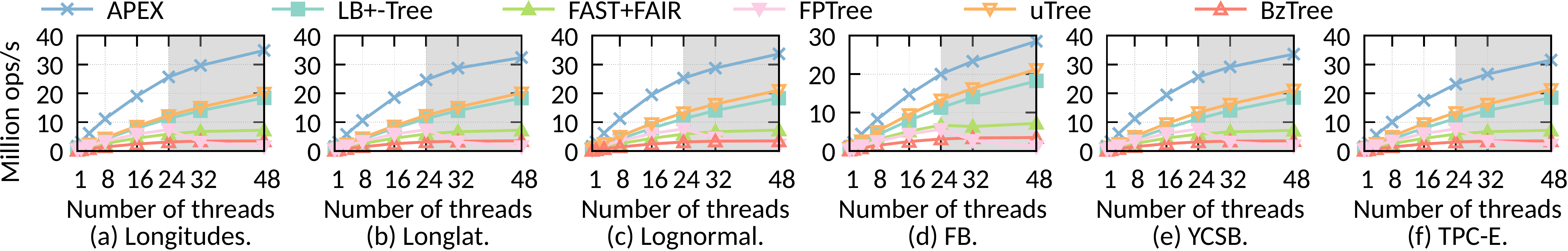}
\caption{Delete scalability.
APEX issues one PM write per delete, and so reduces PM bandwidth consumption and scales better.}
\label{fig:delete-scale}
\end{figure*}

\begin{figure*}[t]
\centering
\includegraphics[width=\textwidth]{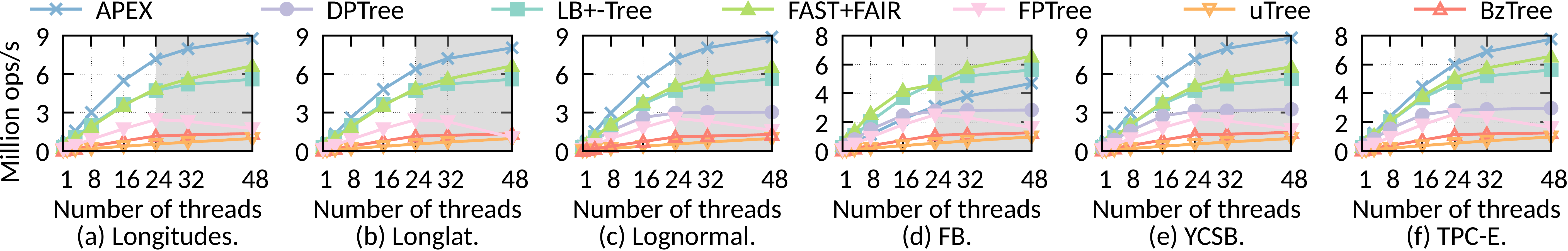}
\caption{Range scan scalability. 
APEX and FAST+FAIR scale well as they maintain records in (nearly) sorted order. 
BzTree/\fptree/\lbtree/\dptree use unsorted nodes, and 
\utree incurs high cache misses when traversing the linked list in PM. 
}
\label{fig:range-scale}
\end{figure*}

\begin{figure}[t]
	\centering
  \includegraphics[width=\columnwidth]{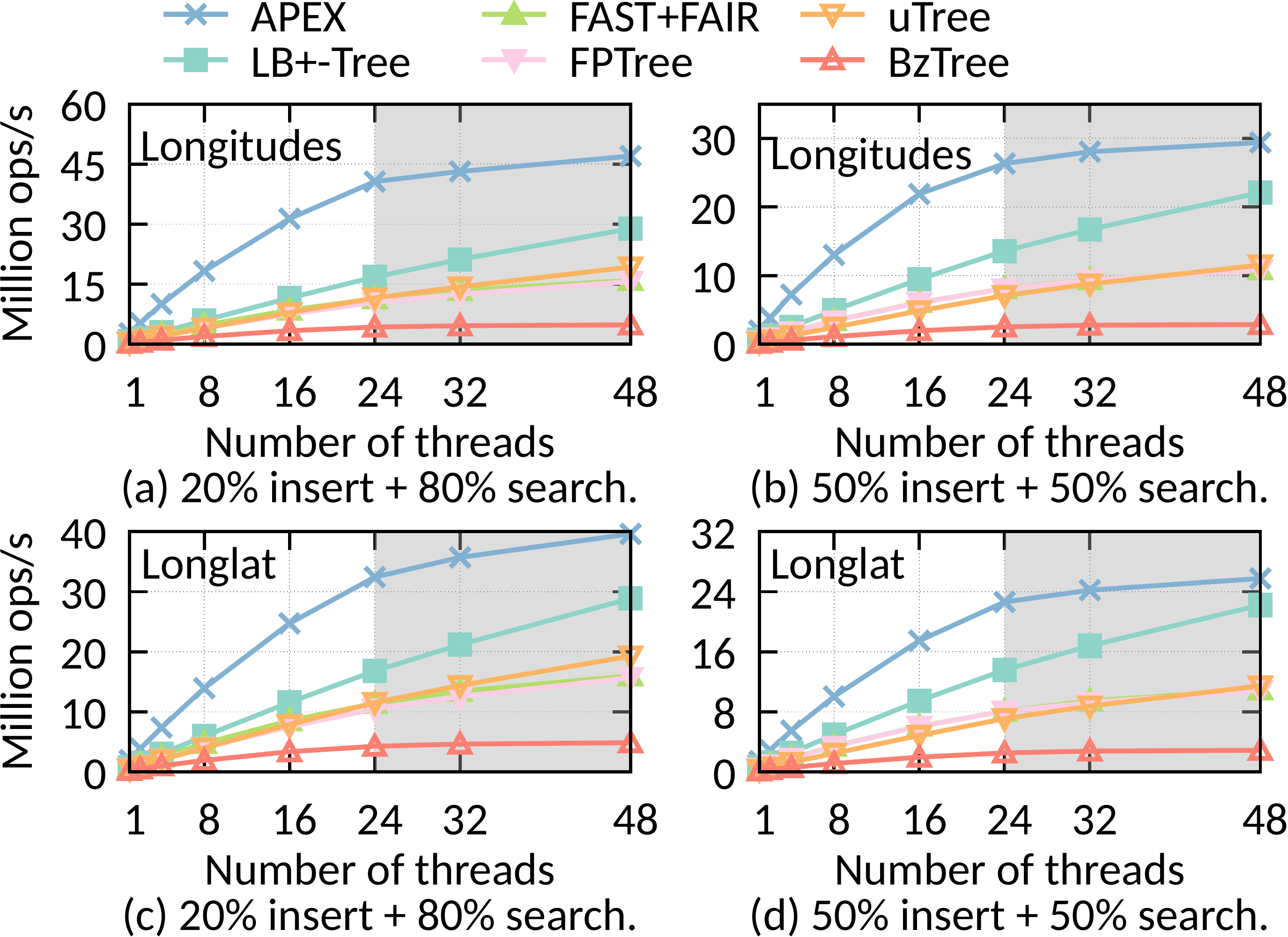}
  \caption{Mixed workload scalability under \lognormal (a--b) and \longlat (c--d).}
	\label{fig:mixed-scale}
\end{figure}

Figure~\ref{fig:single-perf}(b) shows the result for search
operations.  APEX performs up to 7.1$\times$/3.9$\times$/4.1$\times$/3.2$\times$/3.2$\times$/5.8$\times$
higher than BzTree/LB+-Tree/FAST+FAIR/\dptree/\utree/\fptree.  
Notably, APEX's throughput is
$\sim$60\%/37\% higher than that of \lbtree and \dptree under the hardest-to-fit \fb
dataset.  The performance differences across all datasets are due to
how well APEX could fit the data.
For datasets that are easy to fit by linear models, APEX exhibits much
higher search throughput, for two reasons: (1) APEX's index layer
is much smaller and shallow, resulting in much better CPU cache efficiency.  
For example, 
in Table~\ref{tab:bulk-statistic}, the average tree depth on \longitudes is 1.1.  
(2) For easy-to-fit data, most records are stored in PA instead of SA (e.g., overflow ratio is
0.006 on \longitudes), so most lookups only need 
model-based search without probing stashes. For
hard-to-fit data, APEX tree depth can be higher (e.g., 3.13 for \fb), reducing traversal efficiency. 
The probing distance and overflow ratio are also higher, adding more overhead during key lookup.

Search performance also affects update/delete.
APEX only issues one PM write per update/delete, and 
consistently outperforms other indexes in Figures~\ref{fig:single-perf}(c)--(d). 
Thanks to the nearly-sorted order using probe-and-stash,
APEX scans up to
7.6$\times/$1.45$\times$/1.46$\times$/1.83$\times$/ 16.38$\times$/3.1$\times$ faster than
BzTree/\lbtree/FAST+FAIR/\dptree/\utree/ FPTree on five datasets in
Figure~\ref{fig:single-perf}(e). \fb presents the worst case
for APEX, due to long overflow bucket lengths.  Scans incur more
cache misses to traverse the overflow buckets. 
FAST-FAIR does not need sorting. 
\lbtree's sorting overhead is very small for its 
small data nodes (256B). 
BzTree often has to sort much larger (1KB) leaf nodes, adding much
more overhead and so performs 3$\times$ lower than APEX even on \fb.
\utree performs the worst among all indexes since traversing the PM linked list 
incurs lots of cache misses.

\subsection{Scalability with Number of Threads}
\label{sec:multi-thread}

Now we examine how each index scales with an increasing number of threads
under various operations and workloads.

\textbf{Individual Operations.} For inserts
(Figure~\ref{fig:insert-scale}), APEX scales better than others 
before hyperthreading on five datasets and is competitive under \fb.
In most cases, APEX
incurs one XPLine write per insert and its adaptive node sizes 
help reduce synchronization overhead. 
In contrast, FAST+FAIR needs to shift existing records while \utree needs to frequently allocate PM records.
BzTree needs to update much metadata used by itself~\cite{PiBench}.
\lbtree scales well with help of DRAM and small 256B nodes, but with lower 
throughput in most cases.
Note that APEX also scales under \fb, although with lower numbers vs. other workloads. This shows APEX's
concurrency control protocol is lightweight. 
APEX benefits from hyperthreading, but not as significantly as \lbtree since
APEX has much higher single-thread throughput, exhausting PM bandwidth earlier with fewer threads. 
\lbtree uses hyperthreading to better use PM bandwidth, 
whereas APEX use less resource (threads) to fully exploit PM bandwidth and maintain high performance.

APEX scales nearly linearly for lookups before hyperthreading (Figure~\ref{fig:search-scale}).  
It consistently outperforms other
indexes for all datasets benefiting from efficient model-based search
which better utilizes CPU caches. 
Figure~\ref{fig:update-scale}--\ref{fig:range-scale} shows that update, erase and scan scale well as they all benefit from efficient model-based search. 
The one PM-write design in update and delete operation especially make APEX scale better than other indexes because other indexes require more PM writes. 
Since we reduce most unnecessary PM reads and writes in those operations, they could further leverage the hyper-threading to exhaust the remaining bandwidth.

\textbf{Mixed Operations.}  Real-world applications usually involve a
mix of operations.  We evaluate each index with two mixed workloads:
(1) a read-heavy workload that consists of 20\% inserts and 80\%
search, and (2) a balanced workload with 50\% inserts and 50\% search.
Both workload issue 100/90 million inserts from \longlat/\lognormal, leading to a total of 500/450 million and 200/180 million operations,
respectively.
In Figure~\ref{fig:mixed-scale}, we show that APEX scales
the best, and the other indexes show similar trends as before.  As the
workload becomes more write-dominant, the throughput of all indexes
drops. But APEX is still up to 1.69$\times$ better than the next
best performing index, \lbtree.

\textbf{SMO Costs.}
Table~\ref{tab:smo-statistic} lists SMO statistics under insert-only workloads.  
Many more SMOs happen in data nodes than in inner nodes where SMOs are lightweight as shown by the average SMO times.
This justifies our adaptive node size design.
Although we use more locks per node, the overhead is very small (0.02\%-0.04\% time of an SMO) relative to other SMO work (e.g., node allocations).  

\begin{figure}[t]
	\centering
	\includegraphics[width=0.95\columnwidth]{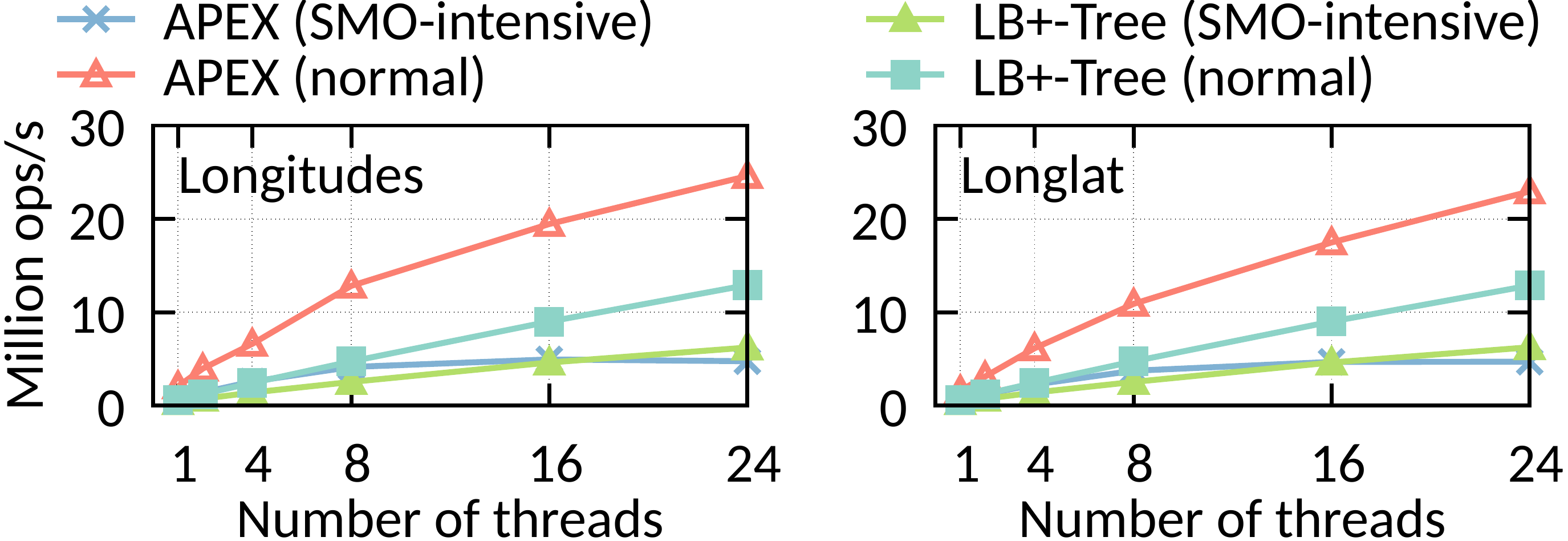}
	\caption{Insert with SMO-intensive vs. normal cases.} 
	\label{fig:smo-overhead}
\end{figure}

We further stress test APEX and \lbtree on SMO-intensive workloads. 
We bulk load APEX/\lbtree with 100 million records with upper density limit 0.9/1.0 (SMO-intensive) and lower density limits 0.5/0.5 (normal) and then run 10 million inserts. 
In Figure~\ref{fig:smo-overhead}, compared to the SMO-intensive cases, APEX and LB+-Tree perform 5.0$\times$/2.1$\times$ better under their ``normal'' cases. 
Under 24 threads, LB+-Tree outperforms APEX by 1.3$\times$ with intensive SMOs as APEX SMO needs to do more work, e.g., model retraining and making decisions based on the cost-models. 
We believe APEX's much higher improvement in the common cases outweighs such degradation on corner cases; we leave more optimizations as future work.

\begin{table}[t]
\centering
\scriptsize
\caption{SMO statistics of insert-only workloads (24 threads).}
\smallskip\noindent
\setlength\tabcolsep{0.8mm}
\begin{tabular}{lrrrrrrr}
\toprule 
\textbf{} & \textbf{\longitudes} & \textbf{\longlat} & \textbf{\lognormal} & \textbf{\fb} & \textbf{\ycsb} & \textbf{\tpce} \\
\midrule  
Inner node expansions&        105 &   2057 &   296 &   5&      1 &     2169 \\
Data node expansions&         15786 & 35277 &  3003 &  73377&  16383 & 139621 \\
Data node splits (sideway)&   1653 &  5327 &   9499 &  440&    1 &     6075 \\
Data node splits (downwards)& 0 &     0 &      193 &   0&      0 &     0 \\
Average inner node SMO time (ms) &    0.11 & 0.09 & 0.66 & 0.94 &  0.01 &  0.03 \\
Average data node SMO time (ms) &   1.74&   0.78 & 2.1 & 0.49 & 1.54 & 0.21 \\
Lock \% during data node SMO& 0.03\%& 0.03\%&  0.03\%& 0.03\%& 0.02\%& 0.04\% \\
\bottomrule 
\end{tabular}
\label{tab:smo-statistic}
\end{table}

\begin{figure}[t]
	\centering
  \includegraphics[width=0.95\columnwidth]{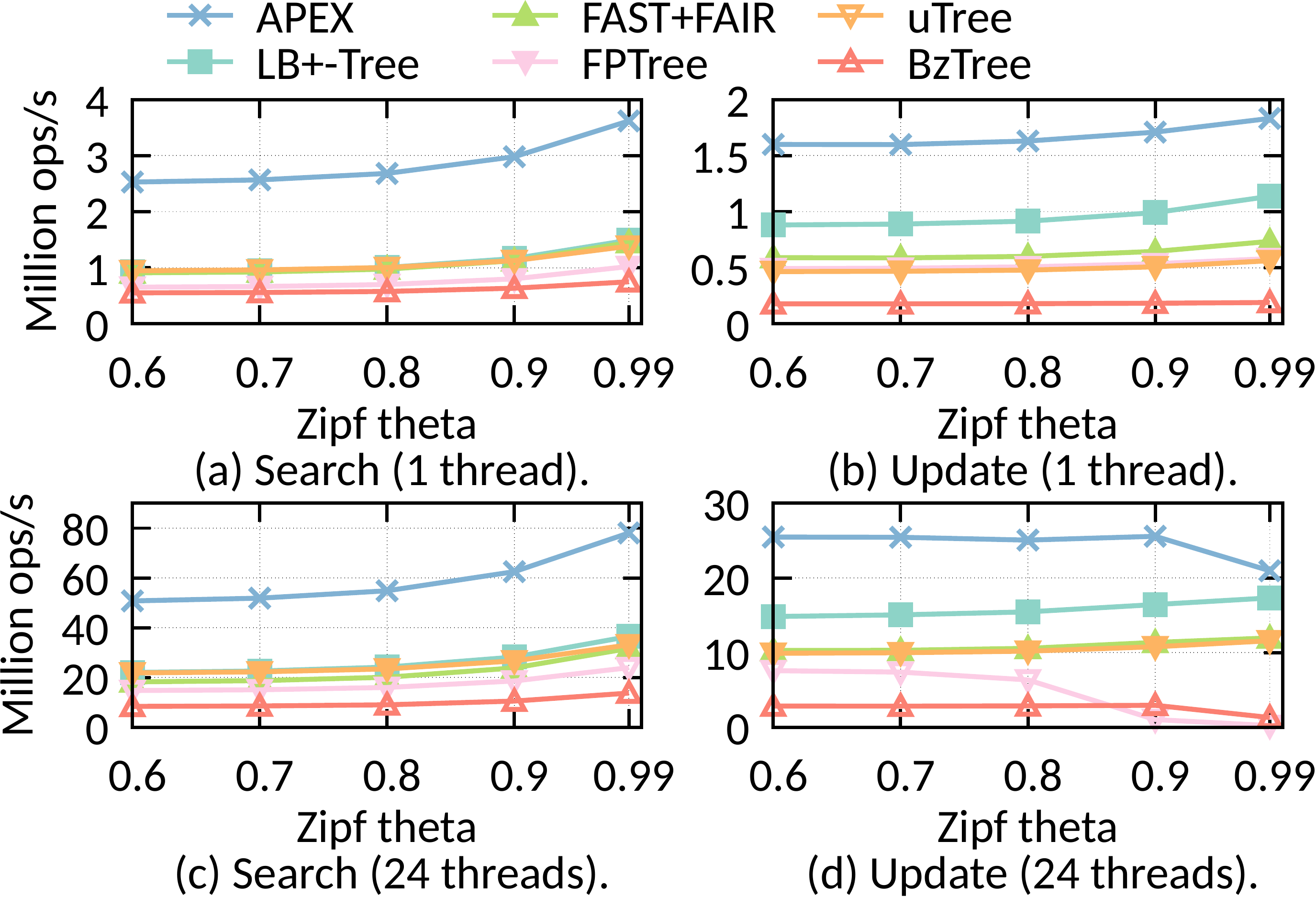}
  \caption{Throughput under varying skewness of Zipfian distribution with one (a--b) and 24 (c--d) threads (\longlat).}
	\label{fig:skew-scale}
\end{figure}

\textbf{Skewed Workloads.} We evaluate search and update operations with 1 and 24 threads under varying skewness. 
As shown in Figures~\ref{fig:skew-scale}(a--b), with a single thread, with higher skewness (theta), all indexes perform better because the accesses are focused on a smaller set of hot keys, better utilizing the CPU cache and less
impacted by PM's high latency. 
The improvement for updates is smaller because they have to flush records to PM.
Under 24 threads in Figures~\ref{fig:skew-scale}(c--d), 
the indexes maintain their relative merits to each other and achieve
higher search throughput because of reduced PM accesses. However, under
very high skew (theta=0.99), contention increases and synchronization
becomes the major bottleneck.

\subsection{Effect of Individual APEX Design Choices}
\label{sec:quantif}
We quantify the impact of APEX's design choices, including node size, probing distance, stash ratio, density bound and accelerators.

\begin{figure}[t]
	\centering
	\includegraphics[width=\columnwidth]{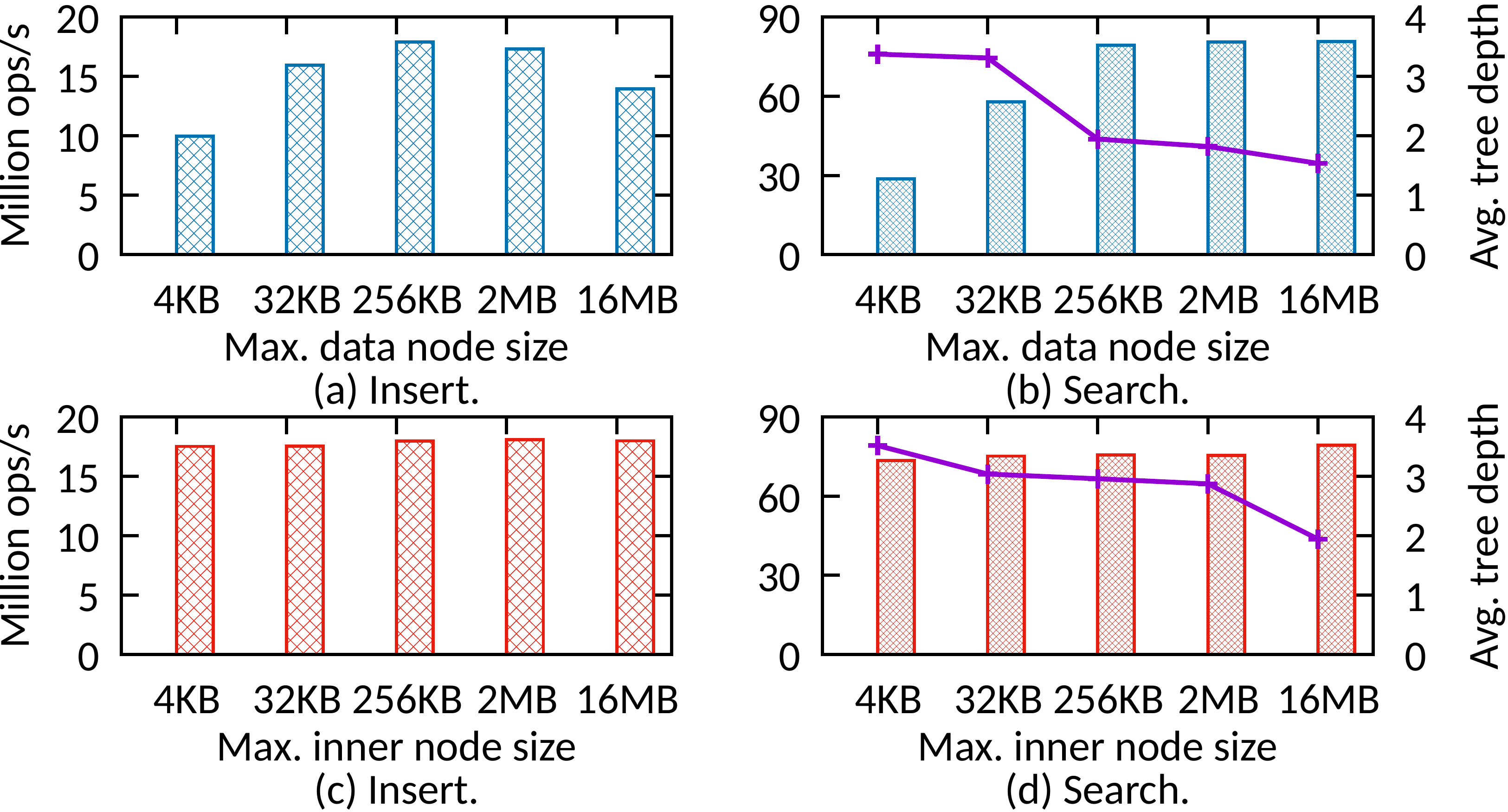}
	\caption{Impact of APEX node sizes (24 threads, \lognormal).} 
	\label{fig:real-node-size}
\end{figure}

\textbf{Maximum Node Size.} 
We start with the impact of maximum node sizes using the easy-to-fit \lognormal as it generates larger nodes close to size limits. 
We first set the maximum inner node size as 16MB and vary maximum data node sizes. 
In Figure~\ref{fig:real-node-size}(a), APEX with 16MB maximum data node size achieves lower performance compared to using maximum 256KB nodes due to higher SMO costs. 
However, using very small data nodes (e.g., 4KB) also leads to lower performance since more inner nodes are needed to index the data, increasing tree depth in Figure~\ref{fig:real-node-size}(b). 
This in turn causes more cache misses during traversal, e.g., 
compared to using maximum 16MB nodes, using 4KB maximum nodes increases the average tree depth from 1.54 to 3.38 with 2.8$\times$ lower search throughput.
Both insert and search performance peak with maximum 256KB data nodes (APEX's default). 
Then we fix the maximum data node size as 256KB and vary inner node maximum sizes. 
In Figure~\ref{fig:real-node-size}(d), although tree depth increases with smaller inner nodes, search performance is barely affected. 
The reason is the inner nodes can all fit in the CPU cache, so 
a deeper tree only needs more computation without much data movement. 
Since SMOs on inner nodes are rare, we also observe little impact on insert performance in Figure~\ref{fig:real-node-size}(c). 
Thus, we set 16MB as the default maximum inner node size to lower tree depth and maintain good search performance.

\begin{figure}[t]
	\centering
	\includegraphics[width=0.95\columnwidth]{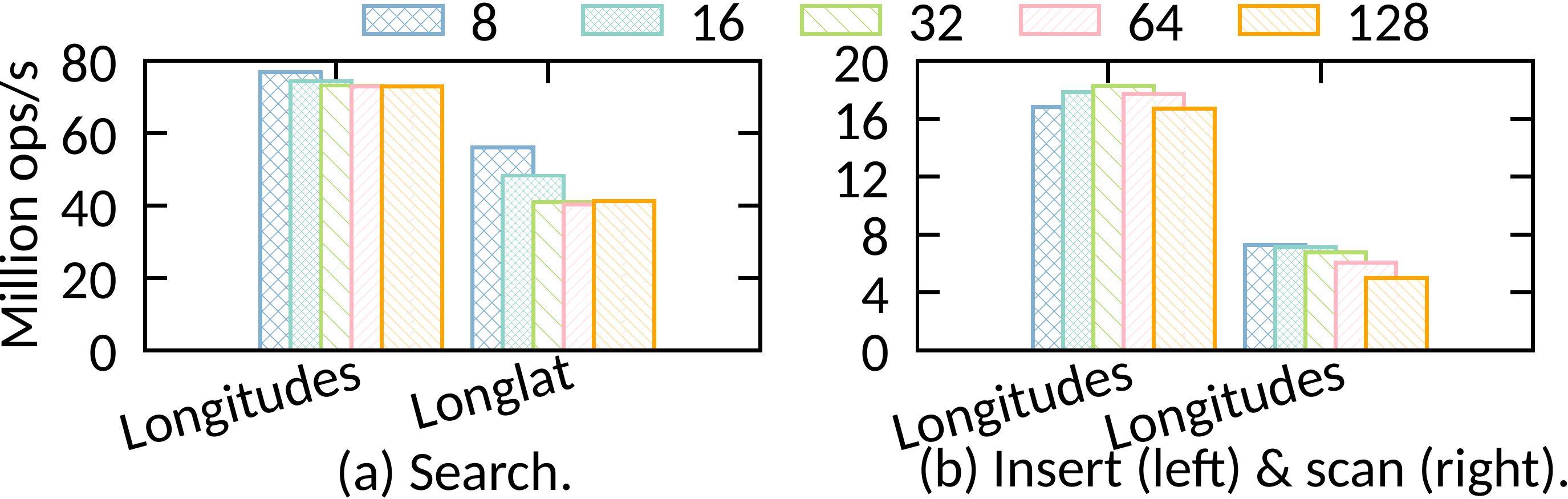}
	\caption{Impact of PA probing distances (24 threads).} 
	\label{fig:real-probing-length}
\end{figure}

\textbf{PA Probing Distance.} 
We study how different bounded (maximum) probing distances $D$ impact performance in Figure~\ref{fig:real-probing-length}. 
For easy-to-fit \longitudes, increasing $D$ barely impacts search since the model fits well. 
But for hard-to-fit \longlat, increasing $D$ from 8 to 128 lowers search performance by 26.5\%. 
Although a larger $D$ reduces SA accesses, 
the average PA probing distance increases (e.g., from 3.6 to 6.5 when $D$ increases from 8 to 128) as collisions are more common in hard-to-fit datasets, pushing records far away from the predicted position.
Using a larger $D$ also mandates more probing in PA before accessing SA.
As Figure~\ref{fig:real-probing-length}(b) shows, insert performance grows by 8\% when $D$ grows from 8 to 32 which more efficiently resolves collisions in PA, thus reducing SA and extended stash accesses. 
However, a large $D$ like 128 reduces insert performance due to the higher cost of uniqueness check in PA.
A higher $D$ also reduces PA's sortedness, lowering scan performance. 
APEX therefore uses $D=16$ to balance search/scan/insert performance.  

\begin{figure}[t]
	\centering
	\includegraphics[width=\columnwidth]{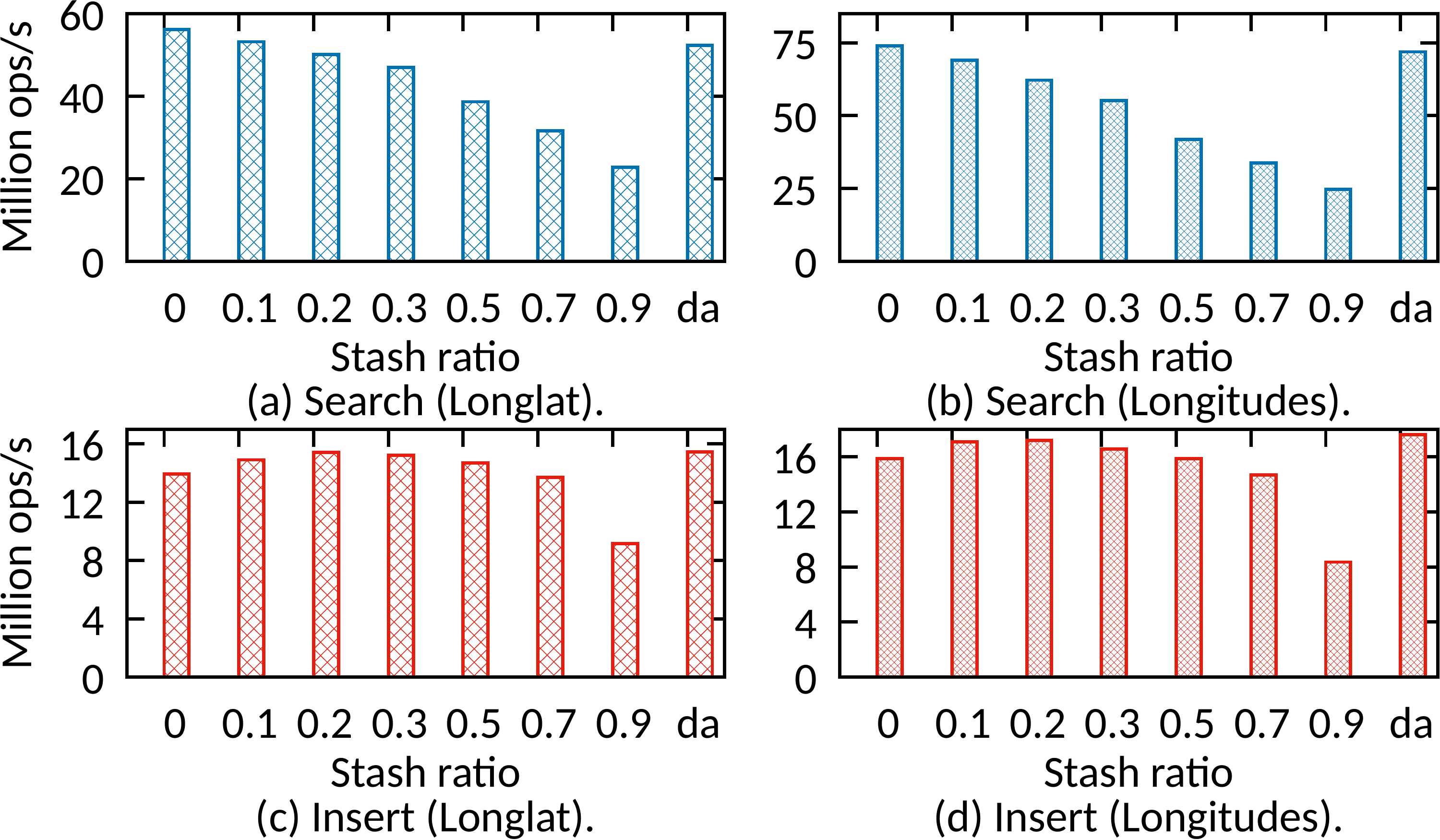}
	\caption{APEX 24-thread throughput under different stash ratios; \texttt{da} indicates APEX's distribution-aware approach.}
	\label{fig:real-stash-ratio}
\end{figure}

\textbf{Stash Ratio.} 
APEX sets SA size based on data distribution and bounds the stash ratio in the [0.05, 0.3] range. 
Now we explore more options by directly setting the stash ratio between 0 and 0.9. 
As Figures~\ref{fig:real-stash-ratio}(a)--(b) shows, search performance drops by 66\%/59\% on \longitudes/\longlat when stash ratio grows from 0 to 0.9. 
Recall that a larger SA leads to a smaller PA, so 
a higher stash ratio leads to more collisions in the smaller PA, routing more lookups in SA/extended stash. 
However, insert performance improves by 8\%/11\% on \longitudes/\longlat when stash ratio increases from 0 to 0.2 in Figures~\ref{fig:real-stash-ratio}(c)--(d) as a relatively larger SA can absorb more inserts and reduce PM allocation costs for extended stash. 
When stash ratio is $>0.3$, insert becomes slower as PA collision overhead cancels out SA's collision-resolving benefits. 
Our distribution-aware approach (\texttt{da}) gives nearly the best performance for both inserts and lookups. 
As Table~\ref{tab:bulk-statistic} shows, the average stash ratios set by \texttt{da} are 0.05/0.118 on \longitudes/\longlat. 
In general, hard-to-fit data like \longlat should have a higher stash ratio to efficiently absorb inserts. 
Easy-to-fit data like \longitudes can use a lower stash ratio to retain the efficiency of model-based search on PA. 
Very low stash ratios will degrade insert performance while high stash ratios (> 0.3) will negatively impact both search and insert performance; 
this leads to our decision to bound the stash ratio in range of [0.05, 0.3]. 

\begin{figure}[t]
	\centering
  \includegraphics[width=\columnwidth]{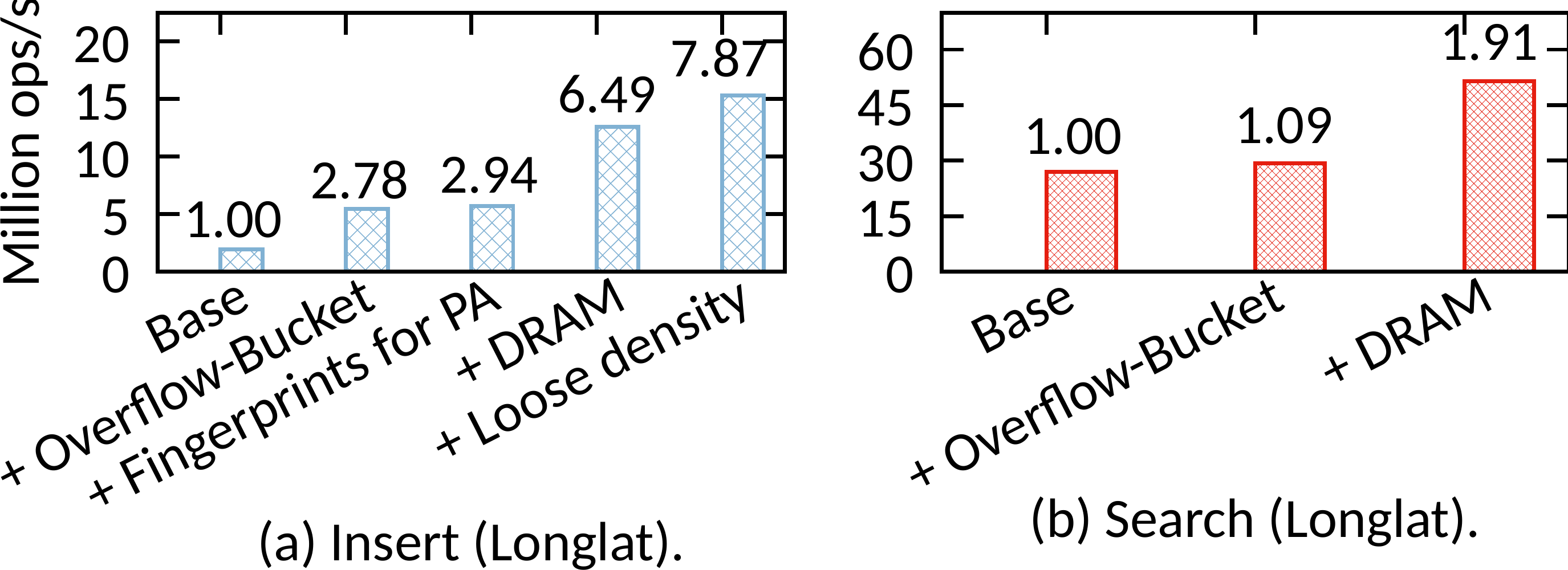}
  \caption{Factor analysis for APEX under 24 threads. Features are added from left to right and are cumulative.}
	\label{fig:ablation}
\end{figure}

\textbf{Effect of Accelerators, DRAM and Density Bound.}
With the recommended parameters fixed, now we explore how each other design choices affect APEX's performance by conducting a factor analysis on one hard-to-fit dataset: \longlat. 
Our results on hardest-to-fit \fb (not shown for space limitation) has a similar trend but larger improvement ratio than \longlat.
We start from the a baseline version that comes with no accelerators, stores the whole tree in PM, and use the tight density bound [0.6,0.8] as ALEX does. 
We then add additional APEX features and observe throughput under 24 threads. 
Figure~\ref{fig:ablation} shows the results. 
APEX with overflow buckets outperforms the baseline version with linear search in SA by 1.09$\times$ by reducing the cost of stash accesses.
The improvement for inserts is 2.78$\times$, which is more significant because the overflow bucket effectively accelerates uniqueness check by avoiding scanning the whole SA. 
PA fingerprints also accelerates uniqueness check during inserts and traversals during update/delete. 
This further improves insert performance by 5\%.
Note that the accelerators added above are still kept in PM, therefore leading to limited improvement.

After placing the accelerators and metadata in DRAM, APEX's search performance increases by 1.76$\times$ because of the reduced PM accesses. 
The increase for inserts is 2.21$\times$. 
Note that the amount of DRAM-resident data used by APEX is small in most cases. 
For example, APEX consumes 0.23/2.14GB of DRAM/PM after loading 100 million records from \longitudes.
It only consumes more DRAM in \fb (0.68/2.29GB of DRAM/PM) since more DRAM-resident overflow buckets are created for stash.
This shows that APEX can leverage PM's high capacity and potentially reduce total system cost, by requiring less or even no DRAM if the user desires. 
\fptree may consume less DRAM~\cite{FPTree} than APEX while APEX always has less DRAM consumption than \utree and is competitive with \lbtree. 
Finally, using a loose density bound ([0.5,0.9]) further improves performance by 1.21$\times$, because 
the loose bound only incurs half of SMOs than using the tight bound, thus issuing less PM writes.

\subsection{Recovery}
\label{sec:recover-time}
We now evaluate how quickly the indexes recover. 
We (1) load a certain number of records, (2) kill
the process to emulate a crash and (3) measure the time needed for the
index to start accepting requests.  Table~\ref{tab:recovery-time}
shows the recovery times for each index on \longlat.  As expected,
the recovery time of \lbtree, \utree and \fptree scales with data size
as their in-DRAM inner nodes need to be rebuilt. 
The main difference between them lies in the leaf-level traversal speed, which is determined by leaf level layout. 
\utree exhibits the longest recovery time as its leaf level is organized in a linked list with one record per node, traversing which incurs many more cache misses than \lbtree's 256B leaf nodes. 
\fptree recovers faster than \utree and \lbtree as its leaf node size is even bigger (1KB) incurring fewer cache misses. 
The other indexes achieve instant recovery (<1s).
At restart, APEX needs to redo/undo in-flight SMOs; such
cases are very rare.  BzTree uses PMwCAS to transparently recover
by scanning a constant-size descriptor pool~\cite{PMwCAS}. 
Both APEX and FAST+FAIR instantly recover with lazy recovery.

\begin{table}[t]
\scriptsize
\centering
\caption{Recovery time (s). APEX can recover instantly with a short warm-up time under 1/24 thread(s) (in parenthesis).} 
\smallskip\noindent
\begin{tabular}{ccccccc}
\toprule 
\textbf{\#Keys} & \textbf{APEX} & \textbf{\lbtree} & \textbf{FAST+FAIR} & \textbf{BzTree} & \textbf{uTree} & \textbf{FPTree}\\
\midrule  
50M& 0.042 (1.94/0.18) & 3.62& 0.042& 0.098 &21.49 &1.63\\
100M& 0.041 (3.74/0.26) & 7.20& 0.041& 0.109 &43.27 &3.26 \\
150M& 0.042 (5.24/0.32) & 10.77& 0.041& 0.097 &65.032 &4.90\\
\bottomrule 
\end{tabular}

\label{tab:recovery-time}
\end{table}

\begin{figure}[t]
	\centering
  \includegraphics[width=0.45\columnwidth]{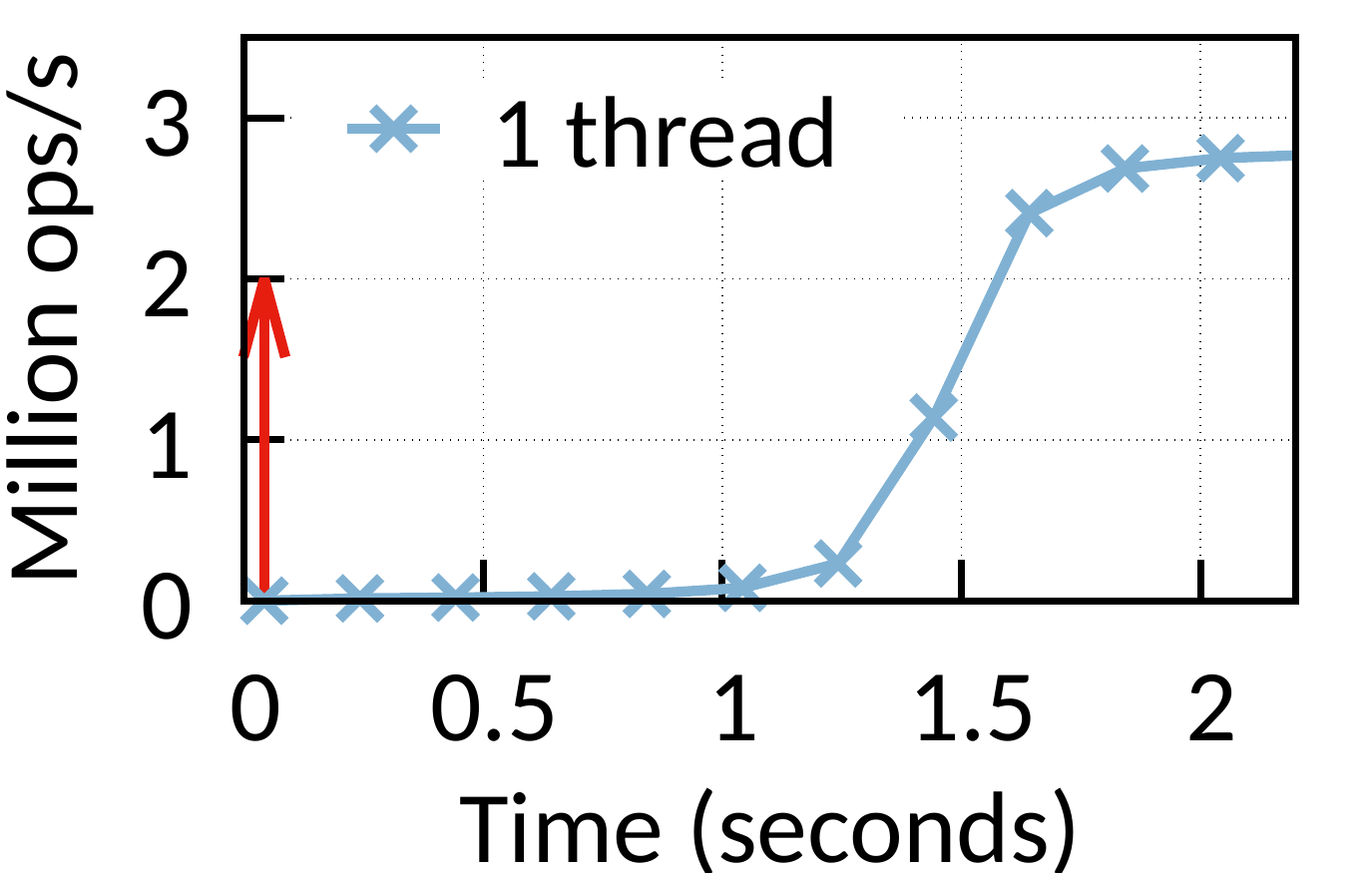}
  \includegraphics[width=0.45\columnwidth]{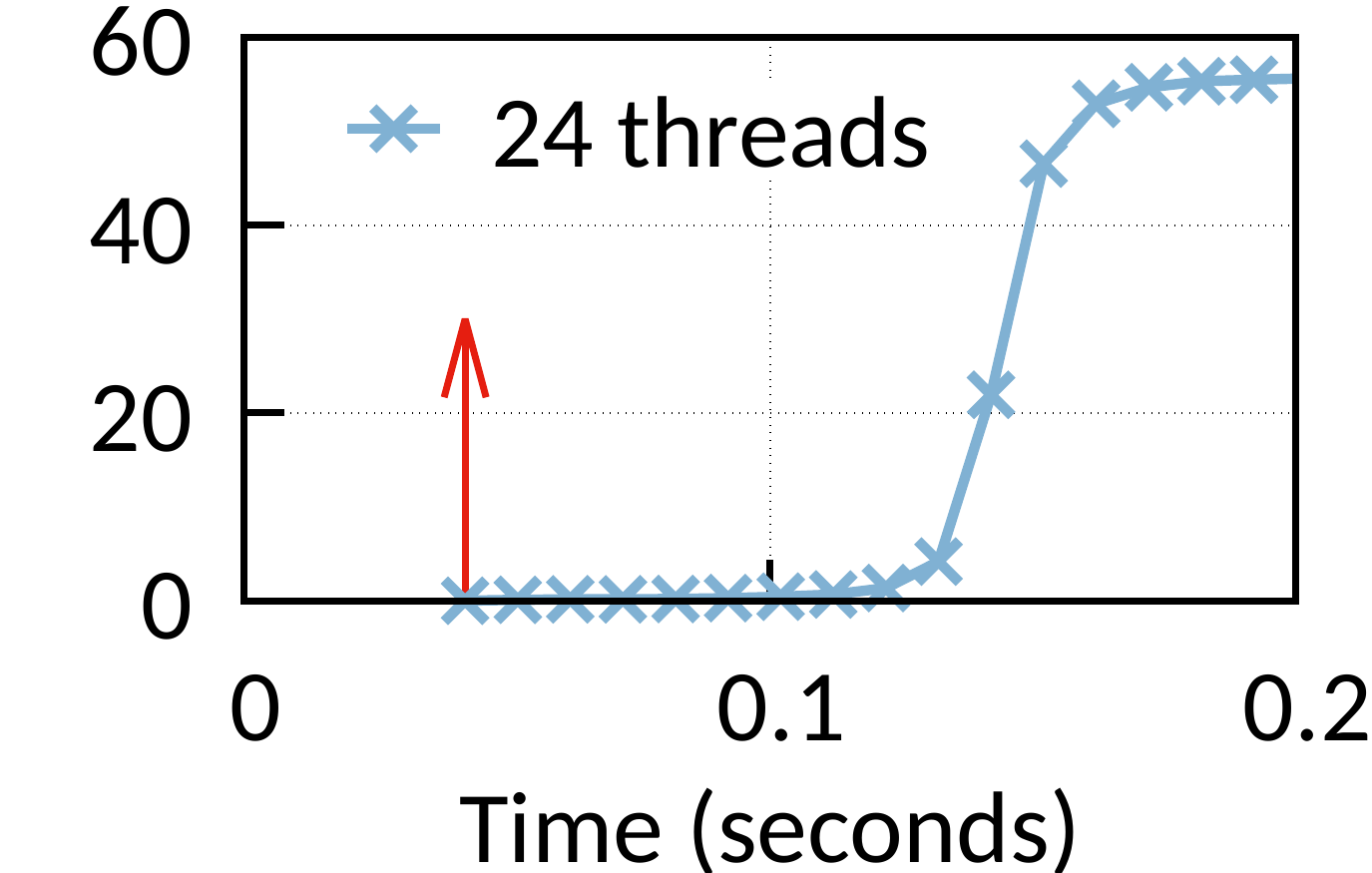}
		\caption{APEX's throughput over time upon restart.} 
	\label{fig:lazy-recovery}
\end{figure}

Since APEX defers the ``real'' recovery work to runtime, we 
evaluate how much warm-up time APEX needs before its
throughput peaks.  We load 50 million records from
\longlat and then kill the process during an insert workload.  After
recovery, we issue lookups and observe throughput. In Figure~\ref{fig:lazy-recovery}, red arrows 
indicate when APEX is ready to accept requests. APEX
initially achieves relatively low throughput: 0.01-0.03 Mops/s with one thread 
thread and 0.1-0.4Mops/s with 24 threads. 
It takes 1.9s/0.15s for throughput peaks with 1/24 thread(s).
Using more threads helps as 
they can recover different data nodes in parallel.
The warm-up time scales with data size, but 
as Table~\ref{tab:recovery-time} shows, it is still faster than \utree and \lbtree and close to \fptree; 
using 24 threads further reduces it to $<1$s. 

\section{Conclusion}
\label{sec:conclusion}

PM offers high performance, cheap persistence and possibility of instant
recovery. Prior work either does not 
exploit the advantages of learned
indexes or PM. 
Yet naively porting a learned index to PM results
in low performance. 
In this paper, we distill several general design principles for
adapting the best of PM and learned indexes. We apply those
principles to the design and implementation of APEX, a concurrent and persistent learned index with instant recovery. Our
in-depth evaluation on Intel Optane DCPMM shows that 
APEX achieves up to $\sim$15$\times$ higher throughput 
compared to recent PM-based indexes, and can instantly recover
in $\sim$42ms. 

\begin{acks}
We thank the anonymous reviewers for their constructive comments.
We also thank Weiran Huang who helped with figure plotting. 
This work is partially supported by an NSERC Discovery Grant, a Canada Foundation for Innovation John R. Evans Leaders Fund, Hong Kong General Research Fund (14200817), Hong Kong AoE/P-404/18, Innovation and Technology Fund (ITS/310/18, ITP/047/19LP) and Centre for Perceptual and Interactive Intelligence (CPII) Limited under the Innovation and Technology Fund, MIT Data Systems and AI Lab (DSAIL), NSF IIS 1900933.
\end{acks}


\bibliographystyle{ACM-Reference-Format}
\bibliography{ref}

\appendix
\newpage

\section*{Appendix}

\section{Complexity Analysis}
\label{sec:complexity}
Now we analyze the complexity of APEX operations. 
Let $n$ and $m$ be the maximum number of slots in inner and data nodes (with PA and SA), respectively. 
Although data nodes may allocate slots in the extended stash, the maximum number of records that can be stored in a data node is fixed ($m\times d_u$ where $d_u$ is the upper bound density) since an SMO will start if the number of records would exceed $m\times d_u$. 
Following ALEX's proof~\cite{ALEX}, traversing from root to leaf takes $\left \lceil log_{n}p \right \rceil$ time, where $p$ is the minimum number of partitions such that when the key space $s$ is divided into $p$ partitions of equal width, every partition contains no more than $m\times d_u$ keys. 
We omit the proof details here and interested readers could refer to ~\cite{ALEX}.

In the best case, inserting or searching a data node takes $O(1)$ time if the model fits well. 
If the predicted PA position for insert is close to a free slot within the bounded probing distance, the record could be inserted in $O(1)$ complexity; 
Future search operation may also directly hit the target record with linear probing in $O(1)$ time.
If the collisions cannot be addressed in the PA, a linear search in the SA may be required to find a free slot, or an extended stash will be allocated. 
The worst-case cost for insert is bounded by $O(m)$ since the extended stash space could not be infinitely allocated given the SMO condition introduced earlier. 
Searching a data node may also need accessing the stash area by linearly traversing the overflow buckets, which again has a bounded cost of $O(m)$.

\begin{figure}[t]
	\centering
	\includegraphics[width=\columnwidth]{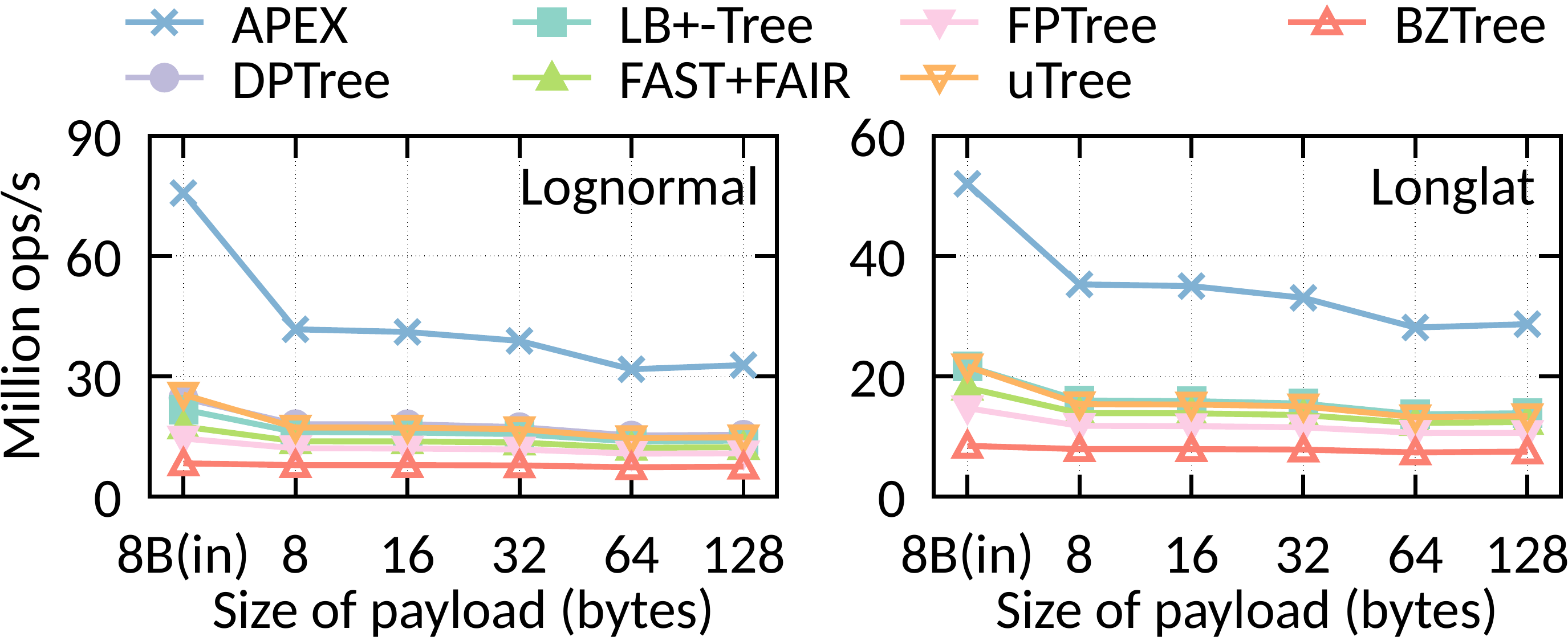}
	\caption{Search throughput under 24 threads with different payload sizes on \lognormal and \longlat.}
\label{fig:payload-size}
\end{figure}

\section{Impact of payload size}
\label{sec:payload}
In this appendix, we study the impact of payload size on the lookup performance of the evaluated trees. 
Using an 8-byte pointer as the payload in index leaf (data) nodes is common in main-memory indexes~\cite{hekaton}. 
The open-source implementations of other trees under comparison~\cite{BzTree,lbtree,Hwang2018,dptree,utree,FPTree} all also assume their payloads are 8-byte pointers. 
Therefore, for all indexes more cache misses are likely to happen when the external payload is accessed. 

To explore how such cache misses impact each index, we conducted an experiment to show how indexes perform when the payload is externally stored with varying sizes (from 8-byte to 128-byte). 
We also include the performance when the payload is an inlined 8-byte value for reference (named \texttt{8B(in)} in Figure~\ref{fig:payload-size}). 
As shown in Figure~\ref{fig:payload-size}, compared to using 8-byte inline payloads, APEX's search throughput drops by 44\%/32\% on \lognormal and \longlat respectively when accessing external payloads. 
Beyond 16-byte payloads, the performance in general drops with bigger payload sizes. 
Our profiling results show that the number of cache misses per search operation of APEX increases from 3.36 to 4.51 on \lognormal (i.e., 34\% more cache misses) when the 8-byte payload is changed from inline-stored to external-stored. 
The numbers for \fptree increases from 9.41 to 10.39 (i.e., 10\% more cache misses), i.e., its performance degradation is not as obvious as APEX's. 
However, even in the condition of external payload accesses (from 8-byte to 128-byte), APEX's search performance still achieves much higher throughput due to the efficient model-based search.

\begin{figure}[t]
	\centering
	\includegraphics[width=\columnwidth]{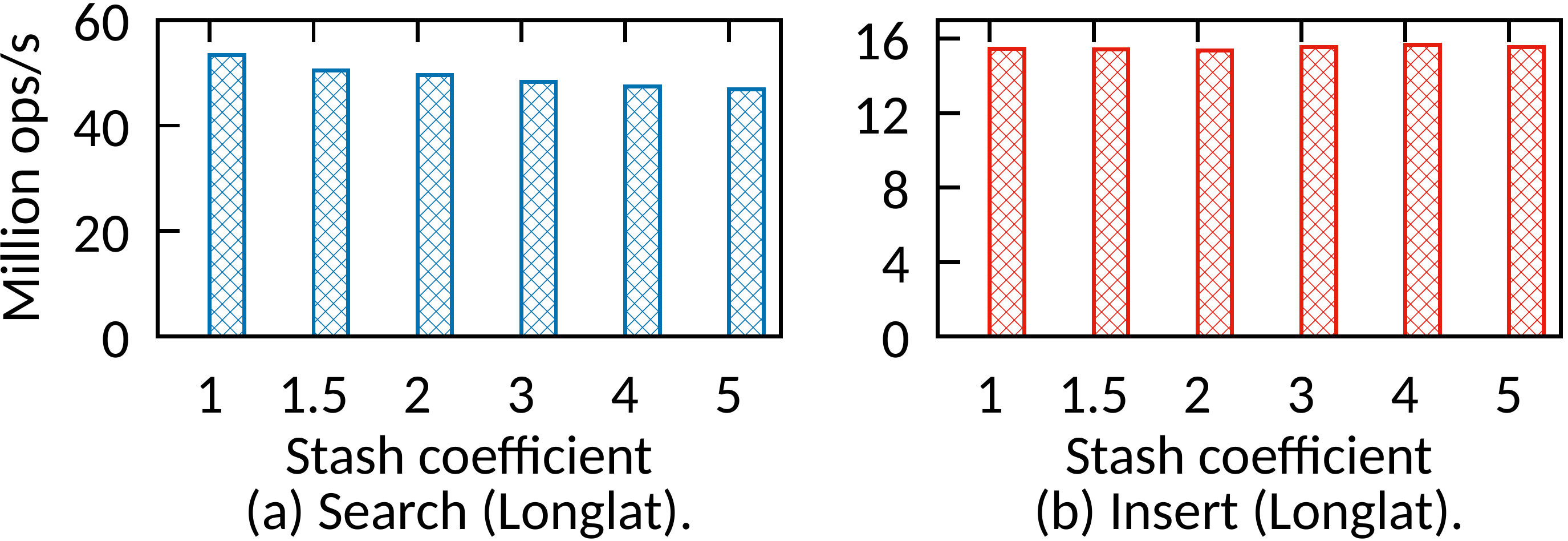}
	\caption{Search (left) and insert (right) throughput of APEX (24 threads) with different stash-coefficients on \longlat.}
	\label{fig:stash-coefficient}
\end{figure}
\section{Impact of stash coefficient}
\label{sec:coefficient}
As stated in Section~\ref{sec:bulk_load}, we set stash ratio $S$ to be a multiple ($n$) of overflow ratio $O$, i.e., $S = max(0.05, min(0.3, n\times O))$ and empirically determined stash coefficient $n$'s value via experiments. 
We show the experimental results in Figure~\ref{fig:stash-coefficient}.
As the figure shows, setting $n$ as 1.5 achieves both good search and insert performance. 
In general, a larger stash coefficient means potentially more keys are stored in the stash areas. 
When varying $n$ from 1 to 5, the average stash ratio increases from 0.09 to 0.22 (i.e., 22\% of the data node's space is allocated to the stash array). 
As the figure shows, the impact is more pronounced for search operations: a larger $n$ (e.g., 3 and 4) means more search operations are routed into stash array but no insert performance improvement. 
We therefore chose 1.5 as the default stash coefficient and overall this gives reasonable performance and is simple to implement/calculate. 
We hope to explore ways to improve on space allocation between primary/stash arrays (e.g., fine-tuning the stash coefficient) to further optimize performance in future work.

\begin{figure}[t]
	\centering
	\includegraphics[width=\columnwidth]{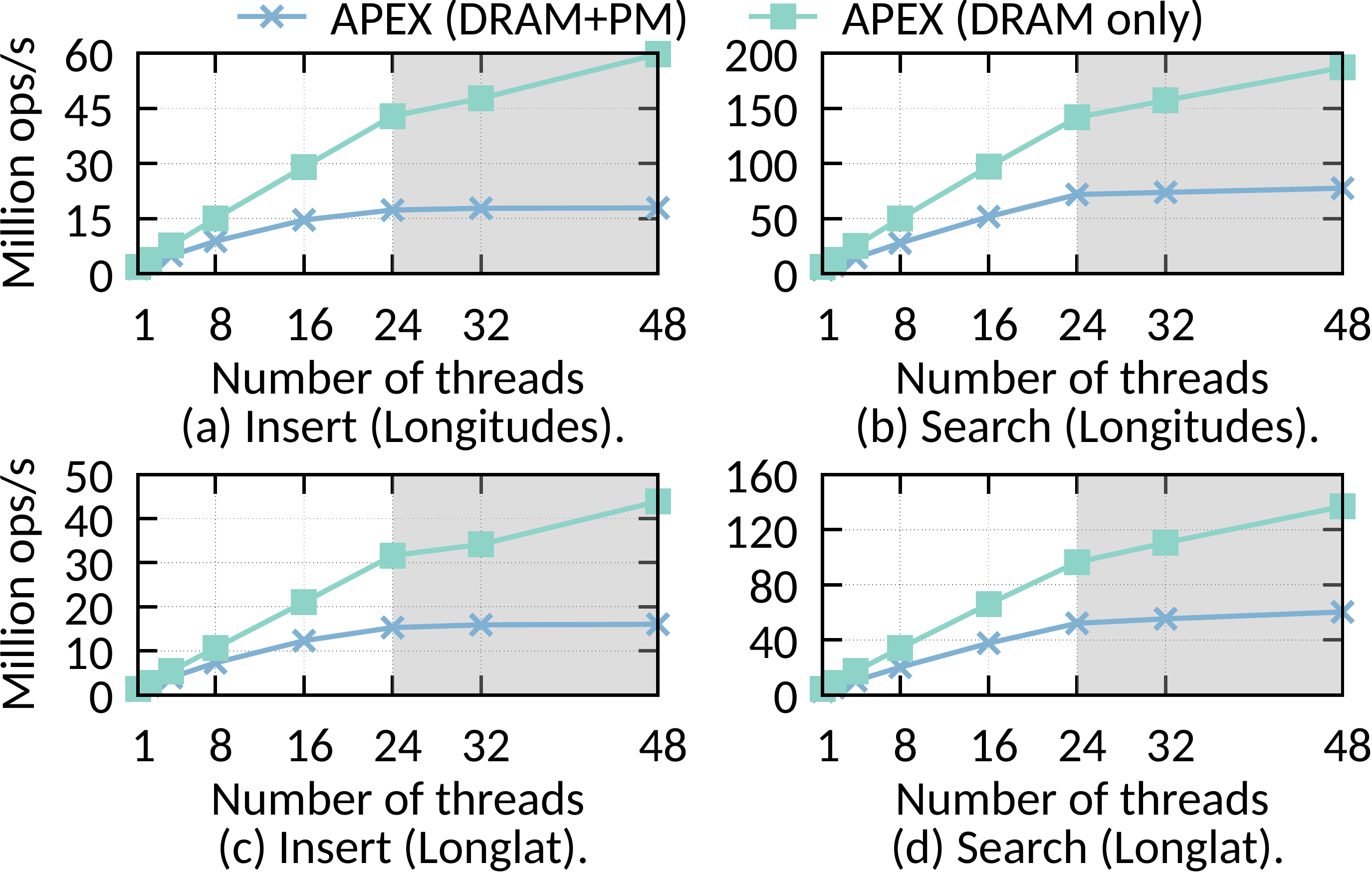}
	\caption{Throughput of APEX (DRAM+PM) and APEX (DRAM-only) on \longitudes (a-b) and \longlat (c-d).}
	\label{fig:pure-dram}
\end{figure}

\section{Comparison with full-DRAM APEX}
\label{sec:full-dram}
We study the impact of PM by comparing with the full-DRAM APEX. 
Specifically, we ran APEX on DRAM by pointing PMDK to use DRAM, same as in prior work~\cite{PiBench} (Lersch et al.). 
As Figure~\ref{fig:pure-dram} shows, the DRAM version performs 2.4$\times$/2.3$\times$ faster on \longitudes and \longlat for search operations. 
The numbers for inserts are 3.3$\times$/2.7$\times$.
APEX on DRAM also achieves near-linear scalability, thanks to DRAM's $\sim$4$\times$ lower latency and $\sim$3-14$\times$ higher bandwidth than PM's.
Also, DRAM's adequate bandwidth allows APEX to benefit from hyperthreading which helps hide CPU stalls. 
Since future PM is expected to have higher bandwidth (e.g., the recent Optane 200 Series already delivers 32\% more bandwidth on average~\cite{optane200}), this experiment also give hints how APEX might perform on newer devices.

\begin{table*}[t]
\centering
\caption{PM and DRAM space consumption (GB). Except for APEX, other indexes only present the memory consumption on \lognormal data set. 
}
\smallskip\noindent
\resizebox{\linewidth}{!}{%
\begin{tabular}{lcccccccccccc}
\toprule 
\textbf{} & \textbf{APEX (\longitudes)}  & \textbf{APEX (\longlat)}  & \textbf{APEX (\lognormal)} & \textbf{APEX (FB)}  & \textbf{APEX (YCSB)}  & \textbf{APEX (TPC-E)} & \textbf{\lbtree} & \textbf{\utree} & \textbf{\dptree} & \textbf{\fptree}  & \textbf{FAST+FAIR} & \textbf{BzTree} \\
\midrule  
PM & 2.139 & 2.159	& 2.152 & 2.29 & 2.138 & 2.244 & 2.649 & 2.98 & 3.03 &  2.542 & 2.47 & 3.46\\
DRAM & 0.232 & 0.311 & 0.226 & 0.676 & 0.227 & 0.265 & 0.265 & 2.396 & 0.055 & 0.055 & 0 & 0\\
\bottomrule 
\end{tabular}}
\label{tab:space}
\end{table*}

\section{Memory (PM/DRAM) consumption}
\label{sec:memory}
For PM/DRAM consumption, we measure the memory consumption of all indexes after bulk loading 100 million records on different datasets ($\sim$1.5GB of raw data), shown in Table~\ref{tab:space}. 
Except for APEX, the memory consumption of other indexes does not vary a lot on different data sets so we only report their memory consumption on \lognormal.
All indexes require extra space than raw data sizes because they reserve empty slots to absorb future inserts. 
Compared to APEX, the other indexes require more PM space because they either store extra metadata in PM (e.g., fingerprints in \fptree) or put the whole tree in PM (e.g., BzTree and FAST+FAIR). 
APEX consumes more DRAM than \fptree and \dptree but its DRAM consumption is always smaller than \utree (up to 13$\times$ smaller) which duplicates the whole index in DRAM. 
In most datasets, APEX has a smaller or similar DRAM consumption as \lbtree.
The only exception is \fb (0.676GB DRAM of APEX vs. 0.265GB of \lbtree), where APEX has to allocate more DRAM-resident overflow buckets for stash due to \fb's extremely non-linear distribution.

\begin{figure}[t]
	\centering
  \includegraphics[width=\columnwidth]{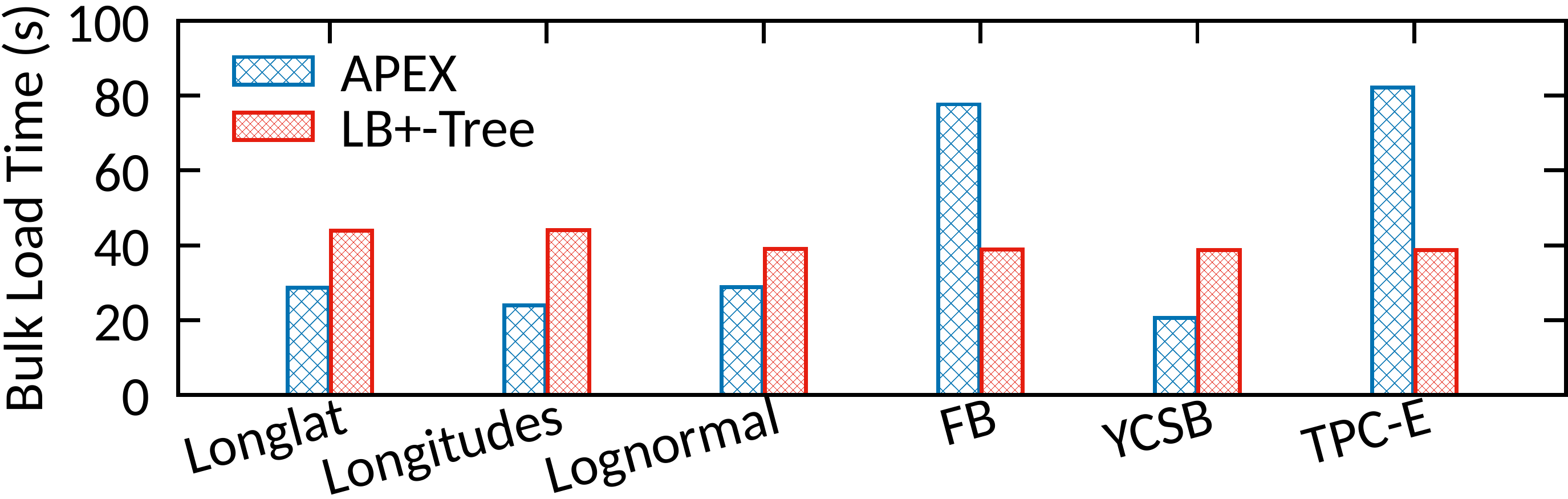}
  \caption{Bulk loading time comparison.} 
	\label{fig:bulk-load}
\end{figure}

\section{Bulk-loading time}
\label{sec:loading-time}
In this section, we study the cost of initial bulk loading. 
We conducted an experiment to show the time for bulk loading 100 million records on various data sets. 
Among the other indexes under comparison, only \lbtree provides a fast bulk loading algorithm while others all require inserting keys from scratch. 
For fair comparison, we test bulk loading time of APEX and \lbtree.
As shown in Figure~\ref{fig:bulk-load}, both trees may perform differently depending on the dataset. 
On average APEX takes 28\% less time. 
In the worst case (\tpce and \fb), APEX is 2.13$\times$ slower than \lbtree because APEX takes more time to search the best node fanout in its fanout space.
We hope to further investigate on improving for such cases in future work.

\end{document}